\documentclass[oneside,onecolumn]{article}

\usepackage{blindtext} 
\usepackage{epsfig}
\usepackage{supertabular,lscape,epsfig}
\usepackage{subfloat}
\usepackage{xcolor}
\usepackage{lineno}
\usepackage{amssymb}
\usepackage{amsmath}
\usepackage{mathtools}
\usepackage{fonttable}
\usepackage{letltxmacro}
\usepackage{subfigure}
\usepackage{cases}
\usepackage{graphicx}
\usepackage{txfonts}
\usepackage[T1]{fontenc}
\usepackage{lipsum}
\usepackage[ps2pdf,%
a4paper=true,%
breaklinks=true,%
colorlinks=true,%
pdfauthor={First Author et al.},%
pdftitle={}%
]{hyperref}
\usepackage[sc]{mathpazo} 
\usepackage[T1]{fontenc} 
\usepackage{microtype} 

\usepackage[english]{babel} 

\usepackage[hmarginratio=1:1,top=32mm,columnsep=20pt]{geometry} 
\usepackage[hang, small,labelfont=bf,up,textfont=it,up]{caption} 
\usepackage{booktabs} 

\usepackage{lettrine} 

\usepackage{enumitem} 
\setlist[itemize]{noitemsep} 

\usepackage{abstract} 

\usepackage{titlesec} 
\renewcommand\thesection{\Roman{section}} 
\renewcommand\thesubsection{\roman{subsection}} 
\titleformat{\section}[block]{\large\scshape\centering}{\thesection.}{1em}{} 
\titleformat{\subsection}[block]{\large}{\thesubsection.}{1em}{} 

\usepackage{fancyhdr} 
\usepackage{titling} 

\usepackage{hyperref} 

\pretitle{\begin{center}\Huge\bfseries} 
\posttitle{\end{center}} 
\title{Empirical Scaling Relations for the Photospheric Magnetic Elements of the Flaring and Non-Flaring Active Regions}
\author{\textsc{M.~A.~Moradhaseli$^1$,~~~M.~Javaherian$^2$,~~~N.~ Fathalian$^3$, H.~Safari$^1$}
\\$^1$Department of Physics, Faculty of Science, University of Zanjan, 45195-313, Zanjan, Iran.
\\$^2$Research Institute for Astronomy and Astrophysics of Maragha (RIAAM), 
\\ University of Maragheh, 55136-553, Maragheh, Iran. 
\\e-mail: javaherian@maragheh.ac.ir
\\$^3$Department of Physics, Payame Noor University (PNU), 19395-3697, Tehran, Iran.
}

\date{\today} 
\renewcommand{\maketitlehookd}{%
\begin{abstract}
Here, we analyzed magnetic elements of the solar active regions (ARs) observed in the line-of-sight magnetograms (the 6173 \AA~Fe \small{I} line) recorded with the Solar Dynamics Observatory (SDO)/Helioseismic and Magnetic Imager (HMI). The Yet Another Feature Tracking Algorithm (\textsf{YAFTA}) was employed to extract the statistical properties of these features (\textit{e.g.} filling factor, magnetic flux, and lifetime) within the areas of $180{\small^{\prime\prime}} \times 180{\small^{\prime\prime}}$ inside the flaring AR (NOAA 12443) and the non-flaring AR (NOAA 12446) for 3 to 5 November 2015 and for 4 to 6 November 2015, respectively. The mean filling factor of polarities was obtained to be about 0.49 for the flaring AR; this value was 0.08 for the non-flaring AR. Time series of the filling factors of the negative and positive polarities for the flaring AR showed anti-correlation (with the Pearson value of -0.80); while for the non-flaring AR, there was the strong positive correlation (with the Pearson value of 0.95). A power-law function was fitted to the frequency distributions of flux ($F$), size ($S$), and lifetime ($T$). Power exponents of the distributions of flux, size, and lifetime for the flaring AR were obtained to be about -2.36, -3.11, and -1.70, respectively; these values of exponents for the non-flaring AR were found to be about -2.53, -3.42, and -1.61, respectively. The code detected a magnetic element with the maximum flux of $23.54\times10^{20}$ Mx, and the maximum value of the size was found to be about 300 Mm$^2$. The most long-lived patch in the flaring AR was belonged to an element with the lifetime of 2208 minutes. It was revealed that $S$, $T$, and $F$ for patches in the flaring AR follow empirical scaling relations as $S \propto F^{0.66}$, $S \propto T^{0.32}$, and $F \propto T^{0.48}$, respectively; For appeared patches in the non-flaring AR, these relations were obtained as $S \propto F^{0.64}$, $S \propto T^{0.23}$ and $F \propto T^{0.37}$, respectively. The comparisons indicated that values of correlations between parameters of $F$ and $T$, and also, $S$ and $T$ for the flaring AR, are more than those of the non-flaring AR, and the quite-Sun (QS). The scaling law relation between the flux growth rate of positive polarities and their size revealed a strong correlation of more than 0.7 in both ARs.
{Sun: flares --- Sun: photosphere --- Sun: magnetic fields --- methods: data analysis --- methods: statistical --- techniques: miscellaneous}
\end{abstract}
}

\begin{document}
\maketitle



\parindent=0.5 cm

\section{Introduction}\label{sect1}
The solar surface is ubiquitously covered by the magnetic elements in which their flux and size distributions follow scale-free behavior (\textit{e.g.,} Parnell \textit{et al.} 2009, Javaherian \textit{et al.} 2017). The magnetic elements emerged in pairs with opposite polarities (Lorrain \textit{et al.} 2006, de Wijn \textit{et al.} 2009). Both of the positive and negative polarities with smallest sizes, known as "patches", are observable in the line-of-sight magnetograms. These magnetic field concentrations are often consistent with appearing bright points in G band and Ca\small{II} H lines (Otsuji \textit{et al.} 2007, de Wijn \textit{et al.} 2009). The size and flux of magnetic patches are ranged from 10$^{14}$ to 10$^{16}$ cm$^2$ and 10$^{17}$ to 10$^{19}$ Mx, respectively (\textit{e.g.,} Stenflo 1973, Hagenaar \textit{et al.} 2003, Priest 2014). The patches' size increases with increasing latitude, and many of patches tend to have appeared as isolated ones in the polar region with high order of magnetic strength about one kG (de Wijn \textit{et al.} 2009).
To understand the underlying process of magnetic elements, many statistical works were done. The power-law behavior of the flux distribution of magnetic features were discussed in Parnell \textit{et al.} (2009). Also, the statistical analyzes of the polarities demonstrated that the magnetic flux growth rate is the power-law function of the magnetic flux (Otsuji \textit{et al.} 2011). By tracking a number of magnetic features in the QS and focusing on their birth (death), it was revealed that for determining the lifetimes of magnetic features, the shredding and dispersal of flux is more dominant processes than the other known processes. The relationships between the network and internetwork solar magnetic flux were studied by Go\v{s}i\'{c} \textit{et al.} (2014). They showed that 14\% of flux originates from internetwork magnetic fields. The analyzes of photospheric plasma motions and polar magnetic patches indicated that magnetic patches' decay is influenced by cancelation and unipolar disappearance with or without fragmentation (Kaithakkal \textit{et al.} 2015). It was found that the rates of appearance and disappearance of magnetic flux in internetwork regions are about $1.2 \times 10^{2}$ Mx cm$^{-2}$ day$^{-1}$ (Go\v{s}i\'{c} \textit{et al.} 2016). Moreover, many articles focused on the statistics and physical properties of photospheric magnetic elements to find their influences on the QS corona (Close \textit{et al.} 2005), network and internetwork areas (Schrijver \& Title 2003, de Wijn \textit{et al.} 2005, Narang \textit{et al.} 2019, Buehler \textit{et al.} 2019), and also, the other solar features such as faculae (Okunev \& Kneer 2004, Kaithakkal \textit{et al.} 2013), coronal mass ejections (CMEs), mini CMEs (Honarbakhsh \textit{et al.} 2016), and filaments (Louis 2019, Liu \textit{et al.} 2019b). Investigating the relation between solar flares as an energetic large-scale and influential phenomena and photospheric magnetic elements as much as predicting solar flares using surface AR patches has been subject of many recent studies (\textit{e.g.,} Conlon \textit{et al.} 2010, Gosain 2012, Wang \textit{et al.} 2012, Burtseva \& Petrie 2013, Jiang \textit{et al.} 2017, Teh 2019, Liu \textit{et al.} 2019a).

Since the evolution of the atmosphere of the Sun is related to the magnetic fields, different high-spatial and -temporal resolution space telescopes such as Michelson Doppler Imager (MDI: Scherrer \textit{et al.} 1995), and Helioseismic and Magnetic Imager (HMI: Schou \textit{et al.} 2012) were established to provide solar photospheric magnetic field magnitudes as magnetogram images. To process the large amount of data in a optimum way, the automatic detection codes are needed. \textit{Curvature} is the name of an algorithm that finds aggregations of pixels with the same polarities which form convex cores around the local extrema. The updated version of this code was represented by Hagenaar \textit{et al.} (1999). Parnell (2002) developed a detection algorithm to group pixels with absolute flux values higher than a given threshold, called \textit{clumping} algorithm. In the other applicable method, named \textit{downhill}, after finding pixels' fluxes with absolute values above lower cuttoff, the structure's peaks are determined, and magnetic features are identified as single ones per local maxima (Welsch \& Longcope 2003). It seems that the downhill and clumping techniques have better segmentation results than the curvature algorithm (DeForest \textit{et al.} 2007). For identifying the magnetic patches in internetwork areas, de Wijn \textit{et al.} (2005) developed a method which works based on convolving an image with a suitable kernel, processing a resultant binary image with spatial and temporal erosion-dilation procedure, and visually inspecting the internetwork bright points candidates after creating mask, temporal averaging in a sequence of magnetograms, smoothing, and intensity thresholding. The other method for the segmentation photospheric magnetic structures was developed by Kestener \textit{et al.} (2010) which works based on 2-D wavelet transform. Otsuji \textit{et al.} (2011) employed the local correlation tracking method to track magnetic elements and determine their morphological evaluation. One of the recent techniques that can track the flux of magnetic features from magnetograms and determines both emergence and cancellation of the magnetic structures is the \textit{magnetic ball-tracking} algorithm (Attie \& Innes 2015). Zender \textit{et al.} (2017) introduced a segmentation method by thresholding on jointing adjacent pixels in a magnetogram. P{\'e}rez-Su{\'a}rez \textit{et al.} (2011) gave a comprehensive review of the automatic detection of solar magnetic features and applications in the space weather.
\begin{table*}[ht!]
\begin{center}
\caption{Characteristics of the flaring AR with NOAA 12443.}
\begin{tabular}{l  c  c  c }
\hline
\\[-2ex]
Day&Latest position of AR& Flares \\[0.6ex]
\hline
\\[-2ex]
3 November 2015 & N06E03 (-50$^{\small\prime\small\prime}$, 31$^{\small\prime\small\prime}$) & C5.5 (18:43 UT), C2.3 (00:39 UT) \\[0.6ex]
4 November 2015 & N06W10 (167$^{\small\prime\small\prime}$, 34$^{\small\prime\small\prime}$) & M3.7 (13:31 UT), C1.4 (03:53 UT) \\[0.6ex]
5 November 2015 & N06W22 (361$^{\small\prime\small\prime}$, 40$^{\small\prime\small\prime}$) & ---\\[0.6ex]
\hline
\end{tabular}
\end{center}
\end{table*}
Javaherian \textit{et al.} (2017) focused on the magnetic features' statistics in the quiet-Sun (QS) over three days. Applying the Yet Another Feature Tracking Algorithm (\textsf{YAFTA}) on the magnetograms, it was found that size, flux, and lifetime distributions of the QS's magnetic features follow the power-law function. By investigating photospheric magnetic elements during the year 2011, they found that the flux and size are strongly correlated. Moreover, the statistical analysis shows that the magnetic features form a self-organized criticality (SOC) system. One may ask whether or not there is any relationship between the magnetic patches in the ARs and the QS. Here, we aim to investigate the features' statistical properties (e.g. filling factor, size, flux, and lifetime) inside the flaring AR and the non-flaring AR. The \textsf{YAFTA} downhill algorithm (DeForest \textit{et al.} 2007) was employed to segregate the magnetic features from the consecutive HMI images. Next, to differentiate scaling laws in the ARs, and also, to compare findings with the results extracted from the QS (Javaherian \textit{et al.} 2017), we focused on the relationships between the physical parameters of the patches. Extracting scaling laws of physical parameters as appeared in the treatment of different phenomena that follow nonlinear process may provide a physical model and an appropriate explanation about the future behavior of the phenomena (\textit{e.g.,} see Aschwanden \textit{et al.} 2008).

The layout is organized in the following steps: we introduce datasets in Section 2. The results are discussed in Section 3. Concluding remarks are given in Section 4.
\begin{figure}
\centerline{\includegraphics[width=1.1\textwidth,clip=]{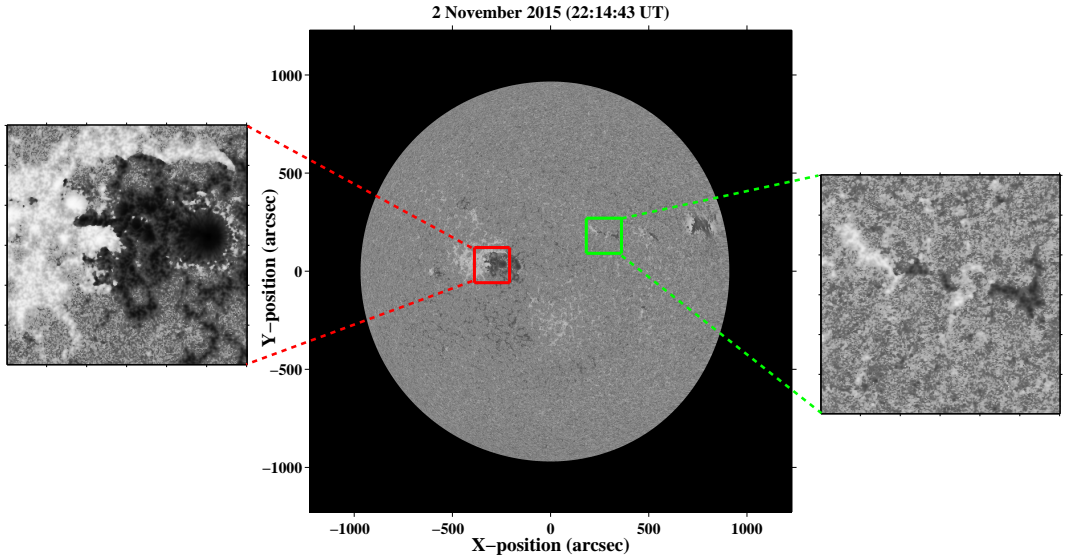}}
\caption{SDO/HMI full-disk magnetogram of the Sun recorded on 2 November 2015. The regions with an area of $180{\small^{\prime\prime}} \times 180{\small^{\prime\prime}}$ (red and green boxes) are selected from all of the sequences for days 3 to 5 November 2015 in the flaring AR (red box), and for days 4 to 6 November 2015 in the non-flaring AR (green box).}
\label{fig1}
\end{figure}
\section{Data Analysis}\label{sec:Data}
Helioseismic and Magnetic Imager (HMI: Schou \textit{et al.} 2012) onboard Solar Dynamics Observatory (SDO) provides full-solar-disk images with spatial and temporal resolutions of about 0.5 arcsec and 45 s, respectively. For our purposes, the magnetogram images with a time lag of 45 s taken at 6173 \AA~ were employed, which recorded from 2 to 6 November 2015. Using the solar software package in Interactive Data Language (IDL) programming, all images were derotated to the reference one when the regions of interest appeared near the solar disk center. We focused on the flaring AR (NOAA 12443), including $C$- and $M$-flares (see Table 1) and the non-flaring AR (NOAA 12446) for three days. We chose square boxes, including the flaring AR (Figure 1, red contour) for days 3 to 5 November 2015, and the non-flaring AR (Figure 1, green contour) for days 4 to 6 November 2015 in the sequence of images. We selected areas within the ARs from solar equatorial regions near the solar disk center (i.e., a region surrounded by longitudes $\pm 17^{\circ}$ around the central meridian, and latitudes limited in $\pm 17^{\circ}$ around the equator) with the size of $180{\small^{\prime\prime}} \times 180{\small^{\prime\prime}}$ to neglect the projection effects (\textit{e.g.,} Alipour \& Safari 2015). In Table 1, the characteristics of the flaring AR and the latest position of AR in both heliographic and heliocentric coordinates were specified (the latest positions of the recorded ARs are valid on 21:35 UT of each day). Then, both constructed data cubes were prepared to be fed to the \textsf{YAFTA} \textit{downhill} algorithm for segregating magnetic features. The label of features is tracked with the \textsf{YAFTA} \textit{match features} algorithm. The two boxes with the size of $180{\small^{\prime\prime}} \times 180{\small^{\prime\prime}}$ were selected from the sequence of derotated HMI images including the flaring AR (Fgiure 1, red box) taken on 3 to 5 November 2015 and the non-flaring AR (Fgiure 1, green box) taken on 4 to 6 November 2015.\\
\begin{figure}
\centerline{\includegraphics[width=1\textwidth,clip=]{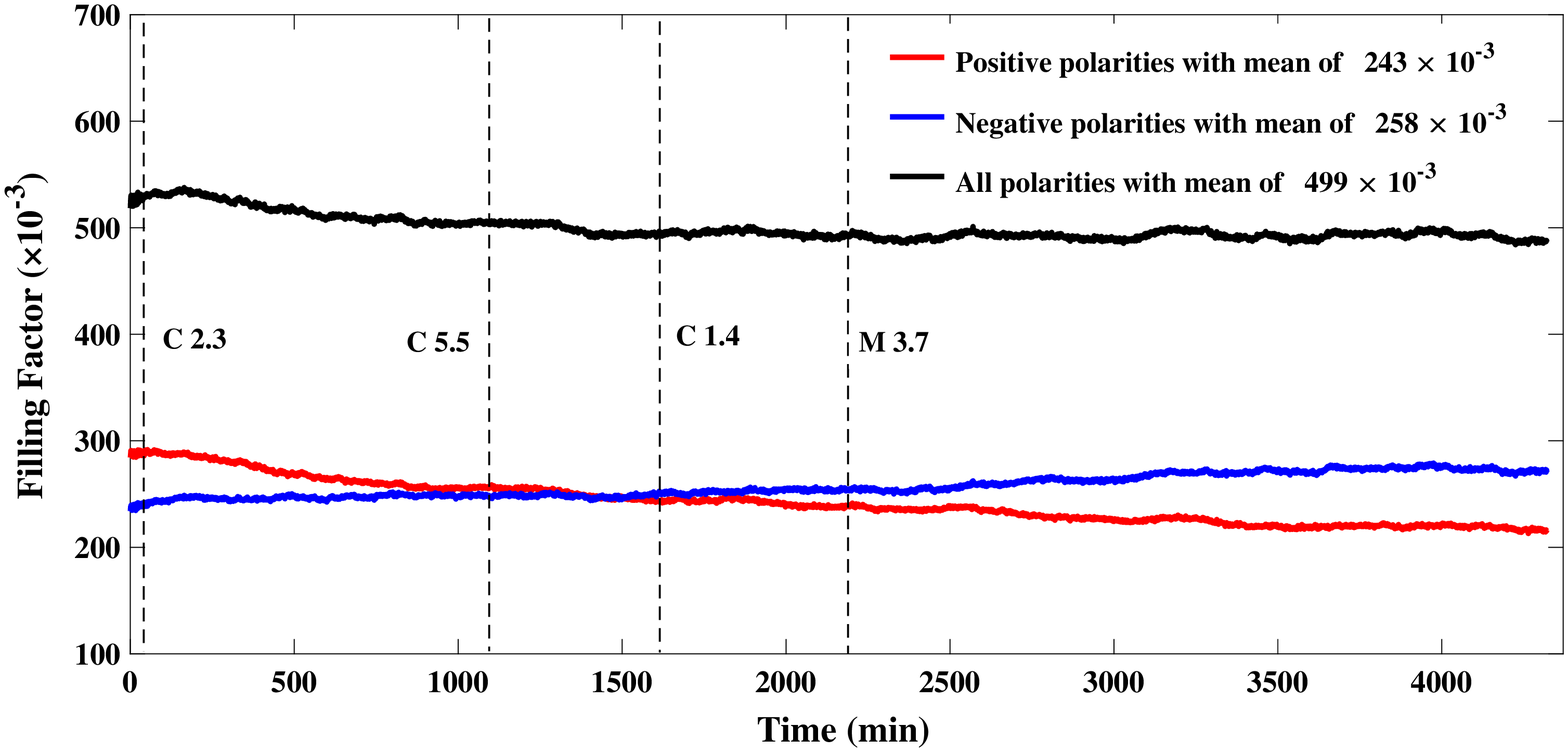}}
\centerline{\includegraphics[width=1\textwidth,clip=]{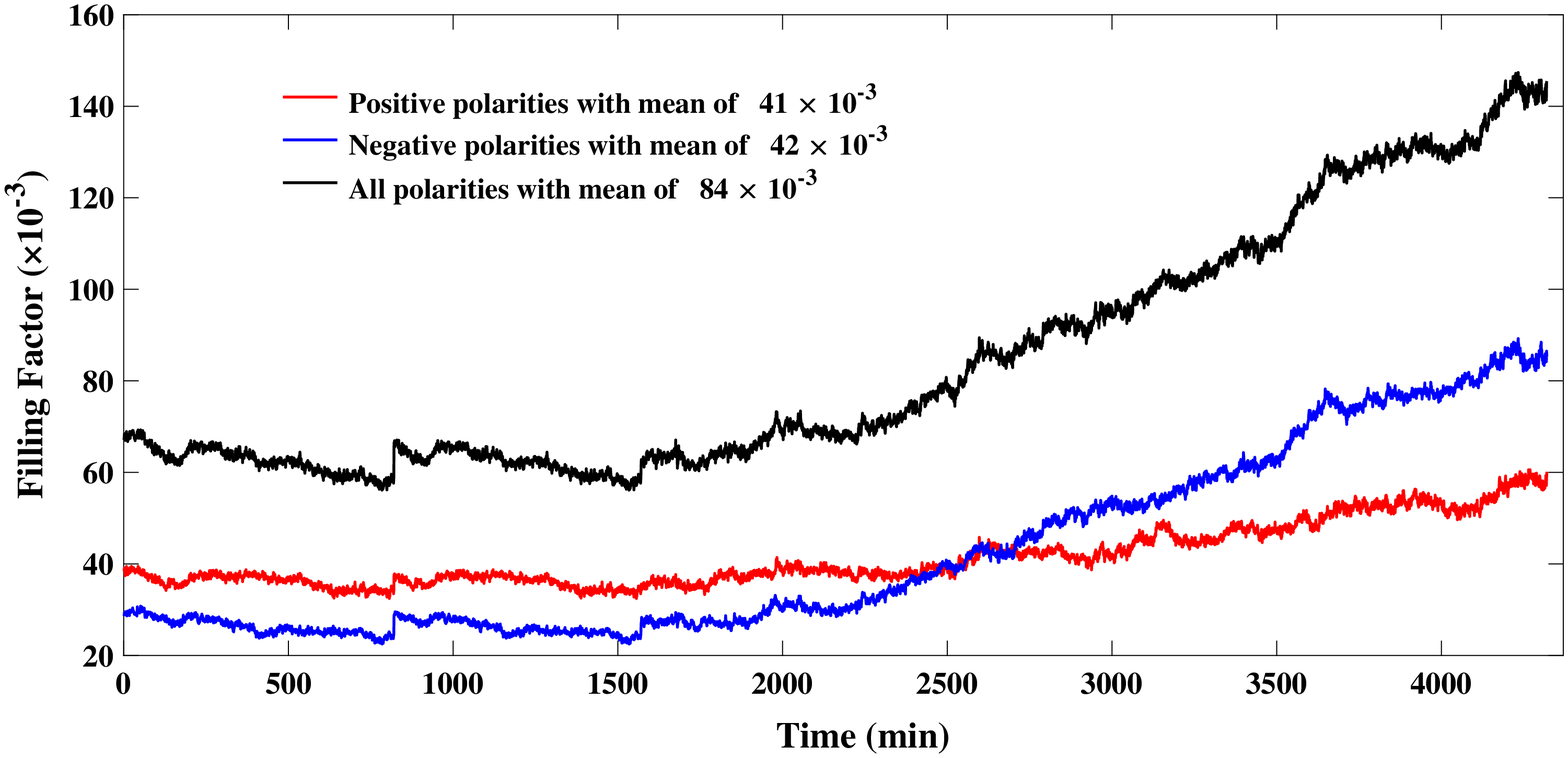}}
\centerline{\includegraphics[width=1\textwidth,clip=]{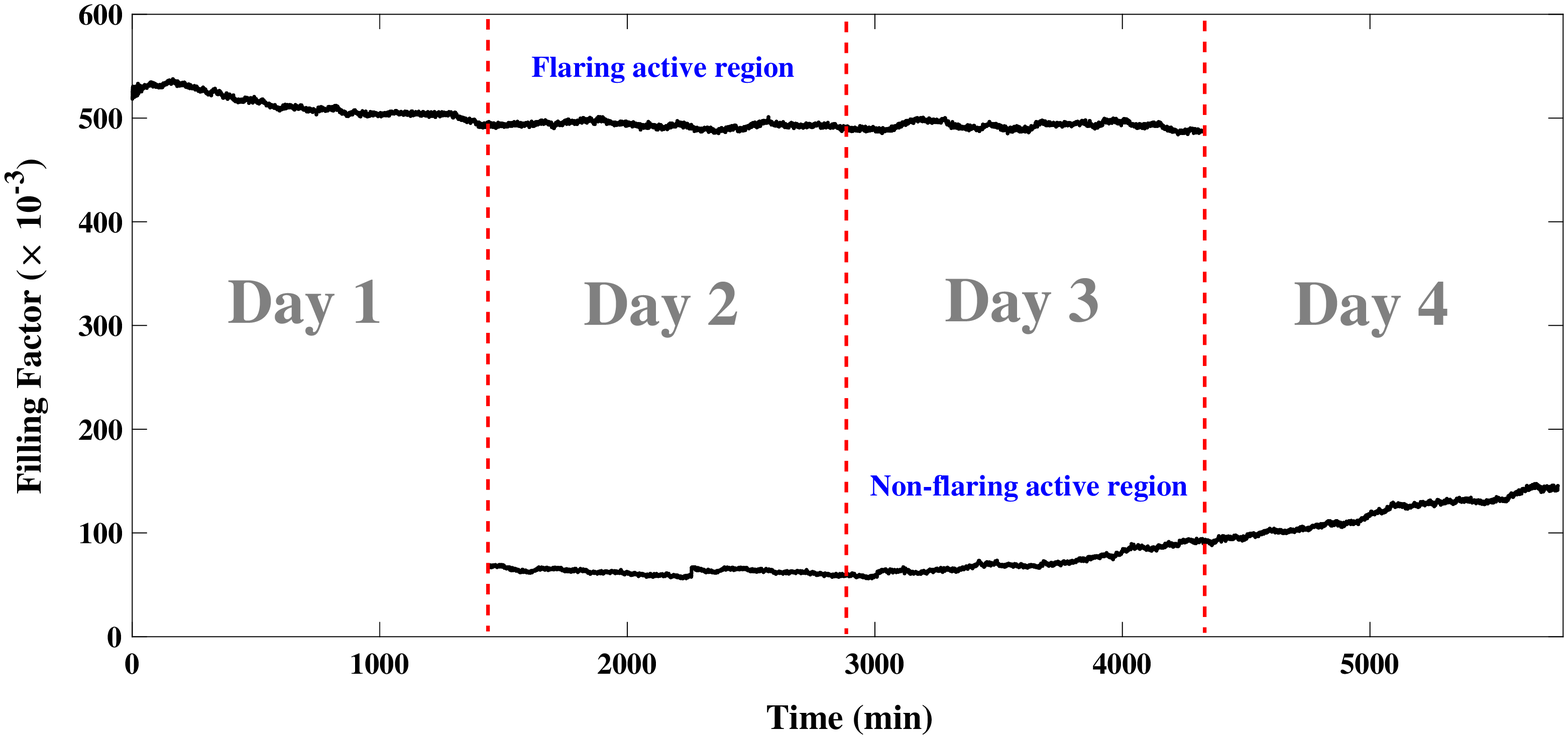}}
\caption{Upper panel: filling factors of positive (red), negative (blue), and all (black) polarities (elements) are plotted for three days (3 -- 5 November 2015) in the flaring AR. The peak times of flaring phase are displayed with dashed vertical lines. Middle panel: filling factors of positive (red), negative (blue), and all (black) polarities are plotted for three days (4 -- 6 November 2015) in the non-flaring AR. The correlated behavior between time-series of the positive and negative polarities is clearly observable. Lower panel: for a comparison, the filling factors of polarities in both ARs have been shown. The area covered by the magnetic elements in the flaring AR is about six times as much as that obtained for the non-flaring AR.}
\label{fig2}
\end{figure}

\section{Results and Discussions}\label{sec:Res}
 The \textsf{YAFTA} code was applied on both constructed data cubes (see Section 2), including three days of data with 5760 sequences. During the sequence of frames in the flaring AR, the code identified 347502 magnetic elements which 250337 of all detected patches survived more than one step (longer than one minute). For the non-flaring AR, among 114063 identified magnetic elements, 70790 of these detected patches survived more than one step. Following the values selected by Parnell \textit{et al.} (2009) and Javaherian \textit{et al.} (2017), the minimum thresholds of the patches area and magnetic field were chosen to be $1.8 \times 1.8$ arcsec$^2$ and 25 Gauss, respectively.

\subsection{Time series of area filling factor, number, and flux of polarities}
To compute the filling factors, the summations of the areas for negative and positive polarities and the whole magnetic features were divided by the size of the selected box in each frame. Figure 2a shows the area filling factors of both positive (red line) and negative (blue line) elements for the flaring AR with the peak times of flaring phase (vertical dashed lines). As it is seen, before the occurrence of the M-type flare (during C-type flaring phase), the lines of the area filling factors of positive and negative elements intersect. In contrast, the summation of filling factors belonging to the both polarities is constant during flaring periods. Figure 2b shows the area filling factors of positive (red line) and negative (blue line) elements for the non-flaring AR. At the beginning of the fourth day of our analyzes (6 November 2015) over the non-flaring AR, the lines of the area filling factors of positive and negative elements intersect, and the whole filling factor of the non-flaring AR grows to reach up about two times of that in the previous day. Figure 2c compares the area filling factors of all magnetic features with their mean values in both ARs. The whole filling factor in the flaring AR ($\sim 0.499$) takes the value approximately six times that of the non-flaring AR ($\sim 0.084$). In Figure 3, the number of the magnetic patches separately for positive (red line) and negative (blue line) elements for the flaring AR (upper panel) and the non-flaring AR (lower panel) are displayed. The mean value of the total number of magnetic patches calculated for the flaring AR ($\sim 1241$) is about six times of the non-flaring one ($\sim 220$). The summations of negative, positive, and total flux of patches with their mean values for both ARs are plotted in Figure 4. The mean value of flux summation in the flaring AR ($\sim 142\times10^{20}$ Mx) is about 14 times of that obtained for the non-flaring AR ($\sim 10\times10^{20}$ Mx).
\subsection{Pearson correlation coefficients between time series}
The Pearson correlation between the filling factors of the positive and negative elements of the flaring AR and the non-flaring AR were obtained to be -0.88 and 0.97, respectively. The Pearson correlation between the number of positive and negative elements of the flaring AR and the non-flaring AR were computed to be -0.64 and 0.19, respectively. As we see in Table 2, the anti-correlated behaviors between the total number of elements and total filling factor in the flaring AR ($\sim -0.28$) and in the non-flaring AR ($\sim -0.20$) represent that the number of the patches increases by decreasing the area filling factor in ARs. On the other hand, the negative and positive fluxes in both ARs have positively correlated behaviors about 0.88 and 0.93, respectively. The measurement of the probability value ($p$-value) test determines the validity of linear correlation between the two time-series. It would take any values in the range of 0 and 1. The $p$-values less than 0.05 imply that the confidence level of obtained correlations is statistically valid (\textit{e.g.} Everitt 2002, Noori \textit{et al.} 2019). As it is seen in Table 2, the probabilities of finding differences for all time-series were computed to be less than 0.1\%.
\begin{figure}
\centerline{\includegraphics[width=1.03\textwidth,clip=]{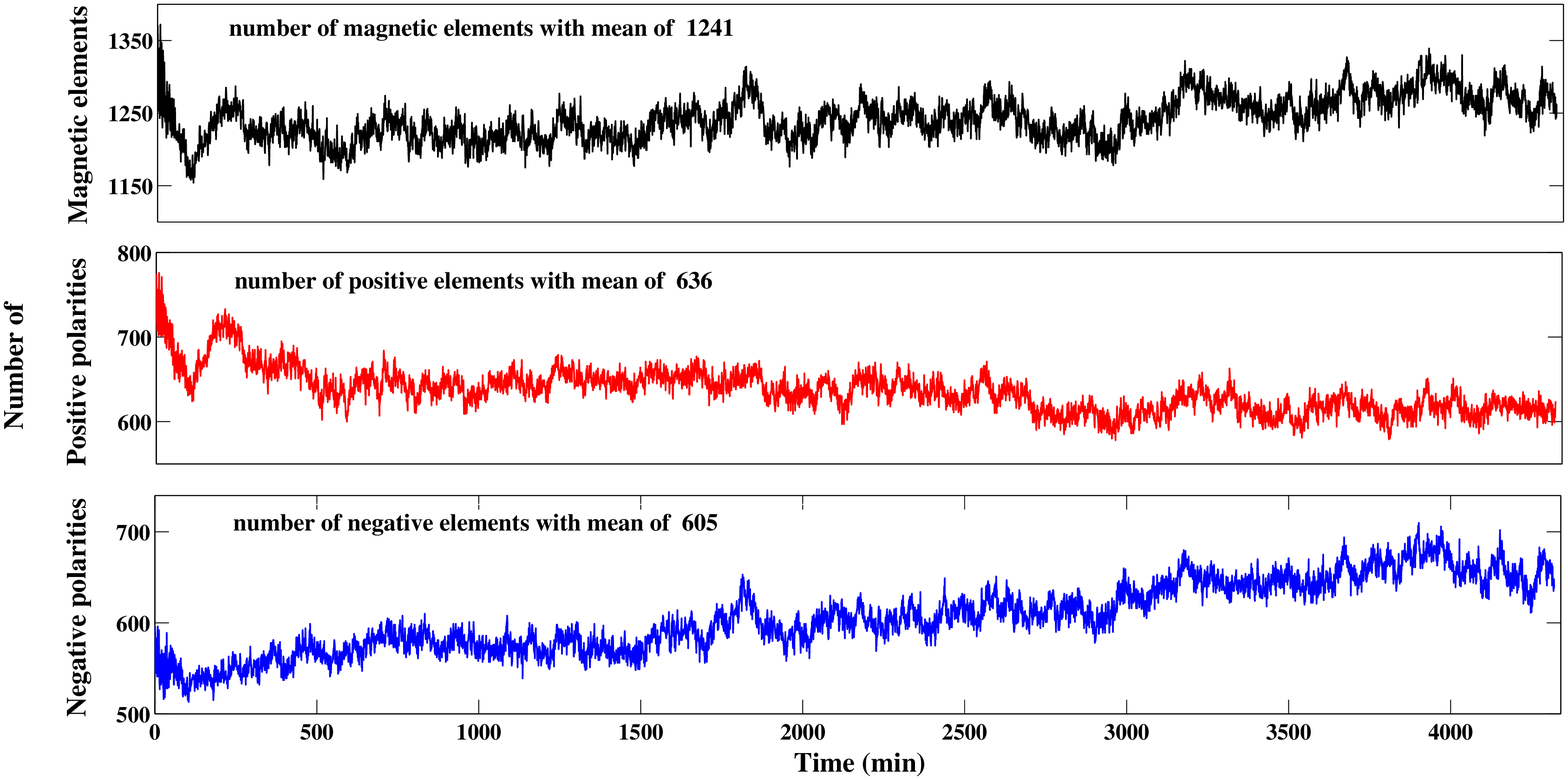}}
\centerline{\includegraphics[width=1.03\textwidth,clip=]{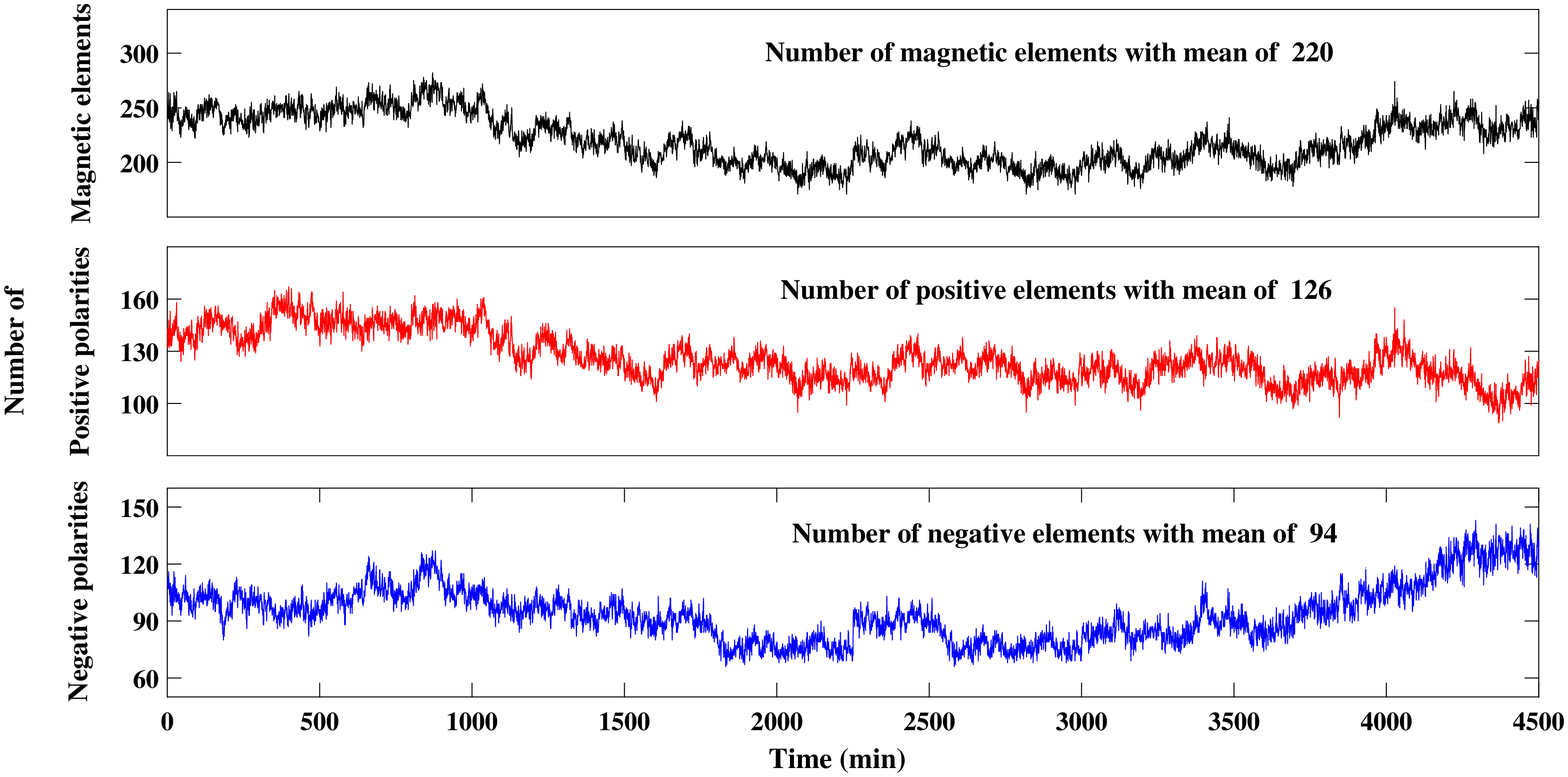}}
\caption{Upper panel: the number of all (black), positive (red), and negative (blue) polarities are plotted for three days (3 -- 5 November 2015) in the flaring AR. It is seen that there are anti-correlated behavior between time series of the number of the positive polarities and the time-series of the number of the negative polarities. Lower panel: the number of all (black), positive (red), and negative (blue) polarities are plotted for three days (4 -- 6 November 2015) in the non-flaring AR. In the same manner as Figure 2, the number of the magnetic elements in the flaring AR is about six times as much as that obtained for the non-flaring AR.}
\label{fig1}
\end{figure}
\begin{table*}[ht!]
\begin{center}
\caption{The Pearson correlations between the parameters of the magnetic elements in the flaring AR (NOAA12443) and in the non-flaring AR (NOAA12446) with $p$-values less than 0.001.}
\begin{tabular}{l  c  c  c }
\hline
\\[-2ex]
Day&Latest position of AR& Flares \\[0.6ex]
\hline
\\[-2ex]
Time series & Correlation value\\[0.6ex]
\hline
\\[-2ex]
  Filling factors of the positive and negative flux in the flaring AR & -0.88 \\[0.6ex]\hline
  \\[-2ex]
  Number of the positive flux and negative flux in the flaring AR & -0.64 \\[0.6ex]\hline
   \\[-2ex]
  Total number of the elements and total filling factor in the flaring AR & -0.28 \\[0.6ex]\hline
   \\[-2ex]
  Positive flux and negative flux in the flaring AR & +0.88 \\[0.6ex]\hline
   \\[-2ex]
  Filling factors of positive and negative flux in the non-flaring AR & +0.97 \\[0.6ex]\hline
   \\[-2ex]
  Number of positive flux and negative flux in the non-flaring AR & +0.19 \\[0.6ex]\hline
   \\[-2ex]
  Total number of elements and total filling factor in the non-flaring AR & -0.20 \\[0.6ex]\hline
   \\[-2ex]
  Positive flux and negative flux in the non-flaring AR &+0.93 \\[0.6ex]\hline
  \\
\end{tabular}
\end{center}
\end{table*}
\subsection{Flux-, size-, and lifetime-distribution of magnetic elements}
The flux-frequency, size-frequency, and lifetime-frequency distributions of all magnetic polarities in the flaring AR are plotted on a log-log scale in Figure 5. We fitted the thresholded power-law function to the distributions, $f(x)$, which is defined as
\begin{equation}\label{thpl}
f(x)=f_0(x_0+x)^{\theta},
\end{equation}
where $x$ is the array of data points, $f_0$ is a normalization constant, $x_0$ is a threshold, and $\theta$ is the exponent which takes the negative value (for more details, see Aschwanden 2015). The genetic algorithm (\textit{e.g.,} Fogel \textit{et al.}1966, Mitchell 1996) was used to find the best model for data points in the procedure of the minimization the chi-square function (For astronomical applications of the genetic algorithm in minimization of chi-square function, the reader can refer to Cant{\'o} \textit{et al.} (2009), and also, Farhang \textit{et al.} (2018)). In our distributions, the parameter $x_0$ was determined to be smaller than the lower bound of the observed data points, $x_{min}$. So, we can use the approximation $(x_0+x)^{\theta} \simeq (x)^{\theta}$ (\textit{i.e.} $(x_0<x_{min}$) for all data points in following calculations (see subsection 3.5). Thus, the power exponents of flux, size, and lifetime were obtained to be $-2.36\pm0.27$ ($R$-square = 0.95), $-3.11\pm0.17$ ($R$-square = 0.96), and $-1.70\pm0.29$ ($R$-square = 0.89), respectively. We found that 96\% of the patches have lifetimes shorter than 100 minutes, while patches with lifetimes longer than 10 hours includes 0.06\% of all detected patches (218 magnetic elements) in the flaring AR. As it is seen in Figure 6, these power-law exponents of the flux, size, and lifetime in the non-flaring AR were obtained to be $-2.53\pm0.20$ ($R$-square = 0.95), $-3.42\pm0.21$ ($R$-square = 0.98), and $-1.61\pm0.19$ ($R$-square = 0.91), respectively. The peak point of flux distributions is about $4\times10^{17}$ Mx in both ARs. As a comparison between the flaring AR and the non-flaring AR, the power exponents of the given frequency distributions of patches show that the magnitude values of slopes in the flux-distribution and size-distribution of the non-flaring AR are greater than those of the flaring AR. However, the magnitude value of slope in the lifetime-distribution of the flaring AR is greater than that of the non-flaring AR. It was revealed that 98\% of patches have lifetimes shorter than 100 minute. There were five patches ($\sim$ 0.004\% of the elements) which have lifetimes longer than 10 hours.
\begin{figure}
\centerline{\includegraphics[width=1.18\textwidth,clip=]{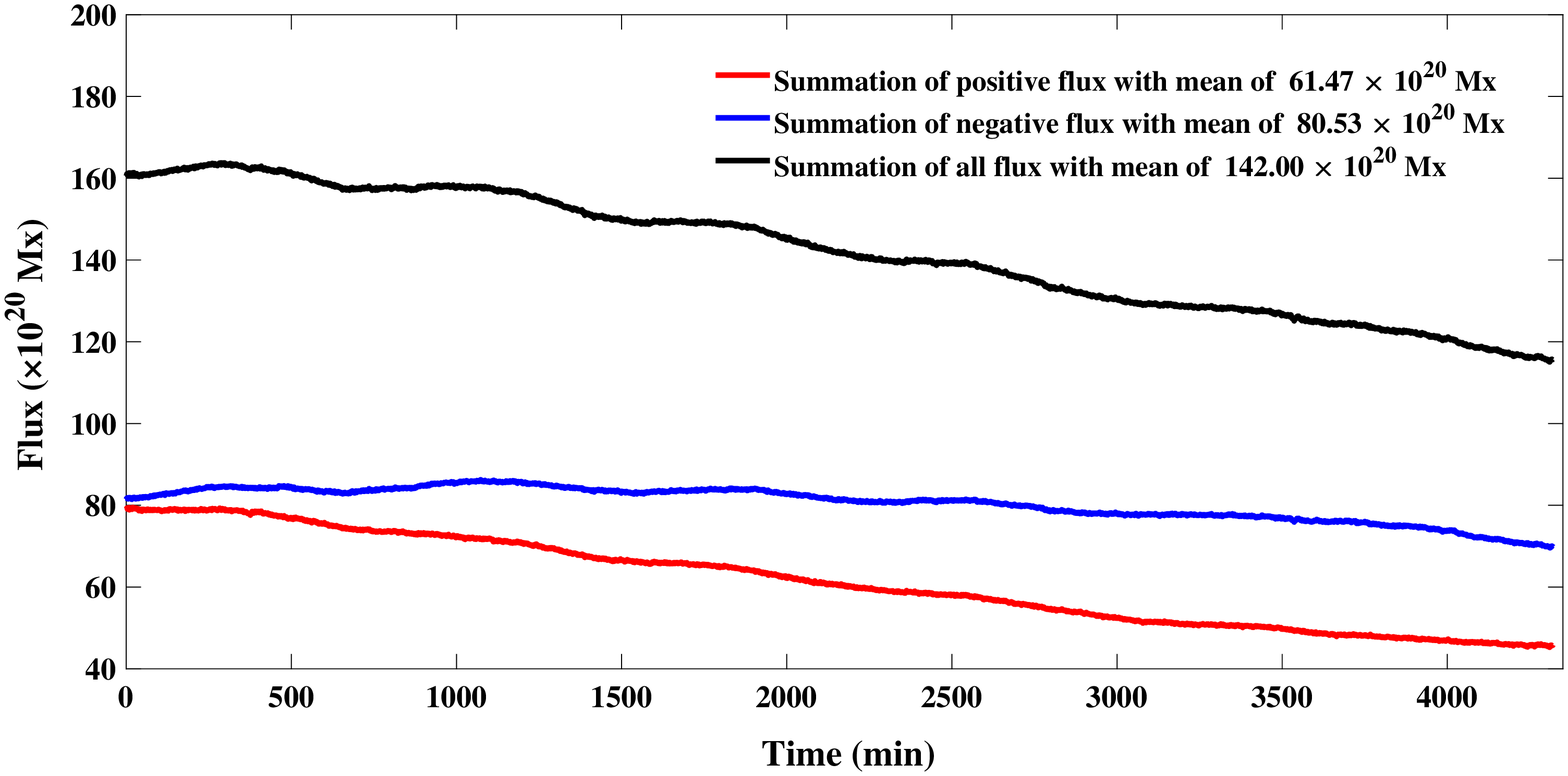}}
\centerline{\includegraphics[width=1.18\textwidth,clip=]{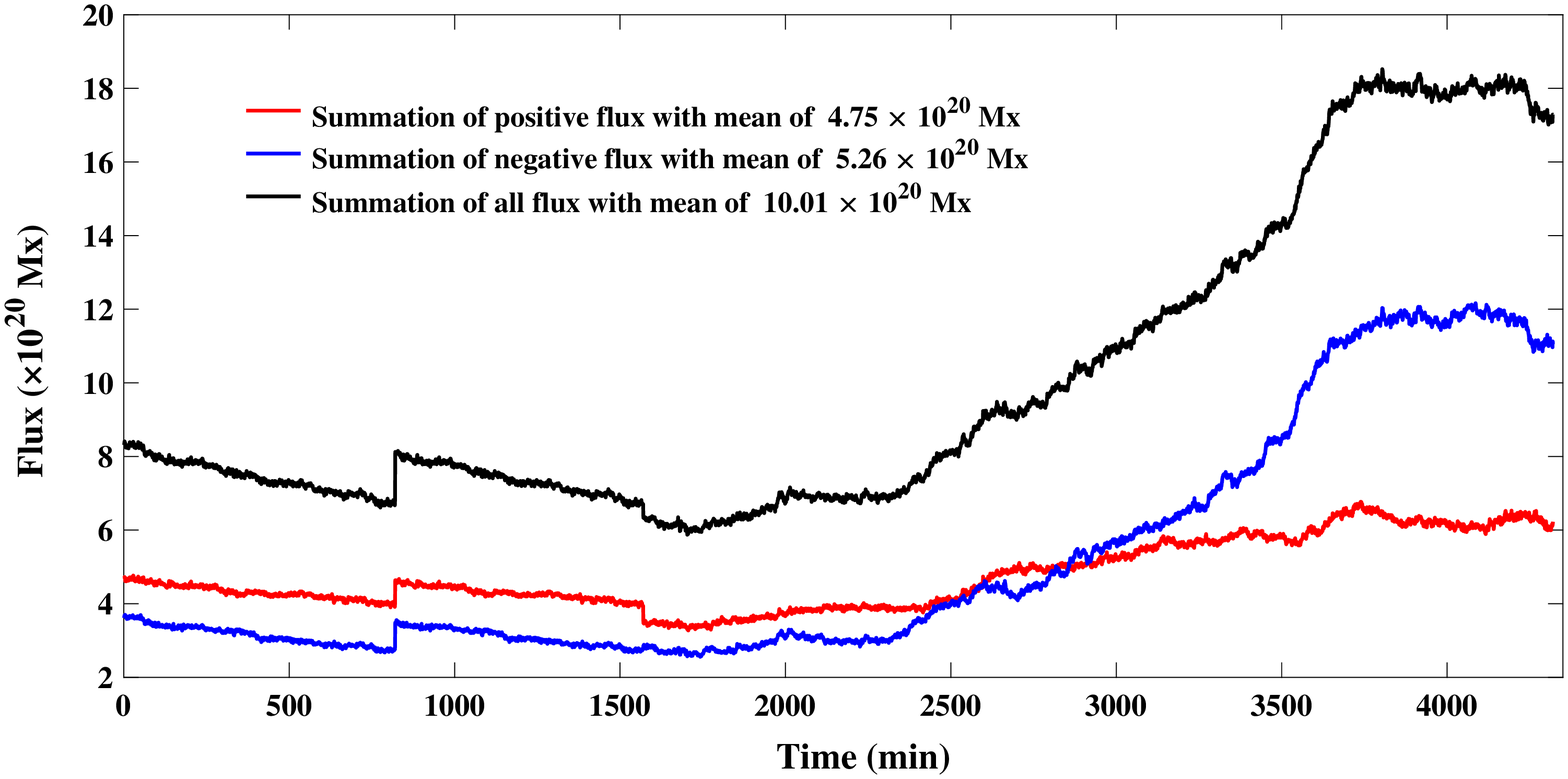}}
\caption{Upper panel: The flux summation of all (black), positive (red), and negative (blue) polarities are plotted for three days (3 -- 5 November 2015) in the flaring AR. Lower panel: The flux summation of all (black), positive (red), and negative (blue) polarities are plotted for three days (4 -- 6 November 2015) in the non-flaring AR. The correlated behavior of positive (red) and negative (blue) polarities is observable in both ARs. The flux summation of all magnetic elements in the flaring AR is about 14 times as much as that obtained for the non-flaring AR.}
\end{figure}
To be sure that the approximation as mentioned earlier gives the accurate exponents, we also employed the maximum likelihood estimator method (MLE: Clauset \textit{et al.} 2009) to apply the whole range of the power-law distributions. The MLE method's task is to estimate the parameters of a model fitted to the frequency distribution by maximizing a likelihood function. For the power-law function with the exponent of $\eta$ and the lower bound of the observed data points $x_{min}$, the following form is used as a model
\begin{equation}\label{powerlaw}
f(x) = \left(\frac{\eta - 1}{x_{min}^{1-\eta}}\right)x^{-\eta}.
\end{equation}
This relation holds for $\eta>1$ and $x_{min}$ can take any values greater than 0. The slope of fit extracted from the MLE is defined as
\begin{equation}\label{powerlawslope}
\eta = 1 + n\left[\sum_{i=1}^n \ln \frac{x_i}{x_{min}}\right]^{-1},
\end{equation}
where $x_i$ denotes the observed values, and $i$ is the numerator started from the first data point up to the last one ($n$) which satisfies the condition $x_i\geq x_{min}$. The first order of standard error value $\sigma$ on $\eta$ is calculated by
\begin{equation}\label{powerlawsigma}
\sigma = \frac{\eta - 1}{\sqrt{n}}.
\end{equation}
The obtained power-law exponents of distributions (-$\eta$), which their standard error ranges are in good agreement with power exponents, are achieved from the thresholded power-law model ($\theta$).
\begin{figure}
\centerline{\includegraphics[width=1\textwidth,clip=]{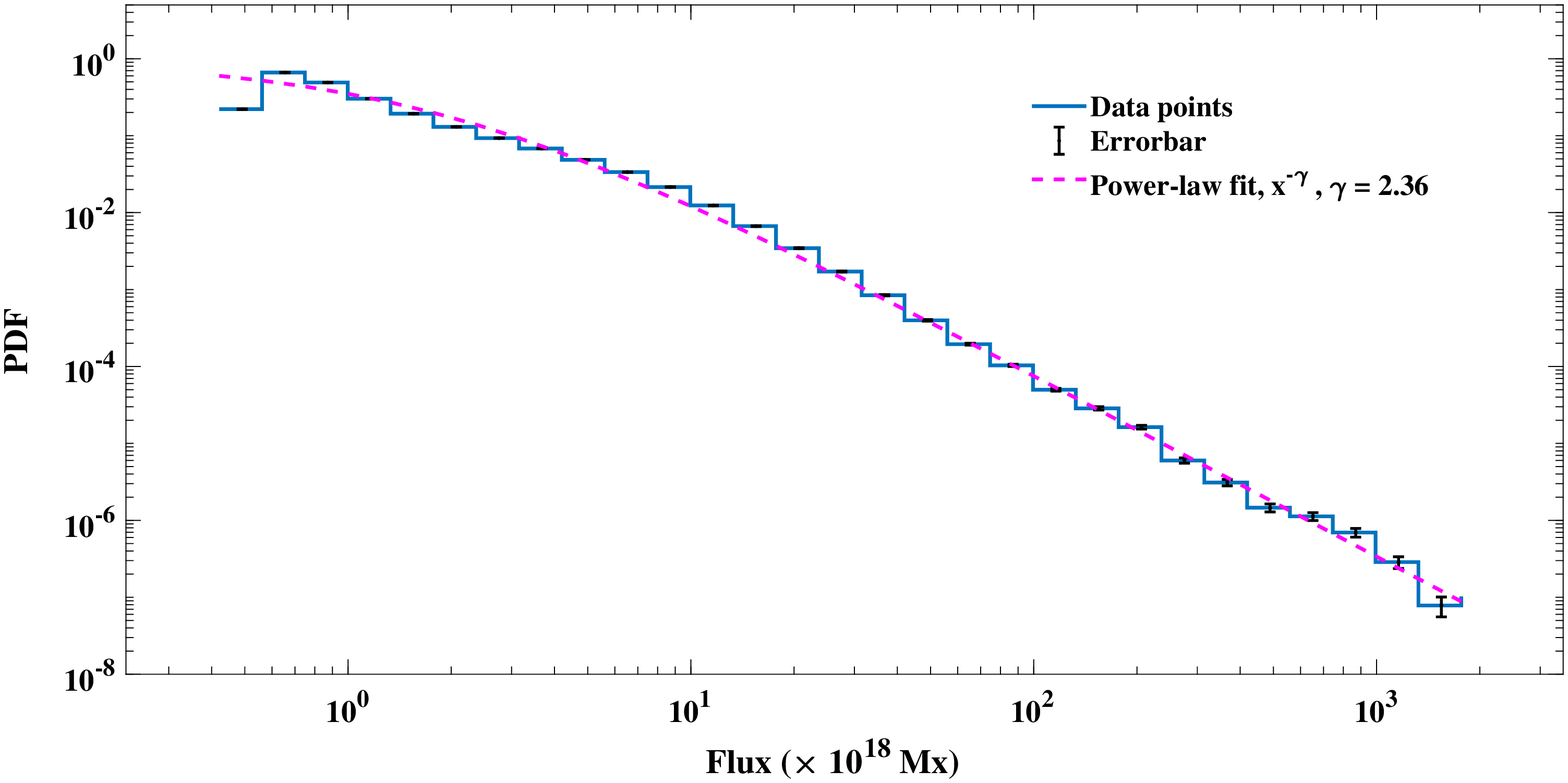}}
\centerline{\includegraphics[width=1\textwidth,clip=]{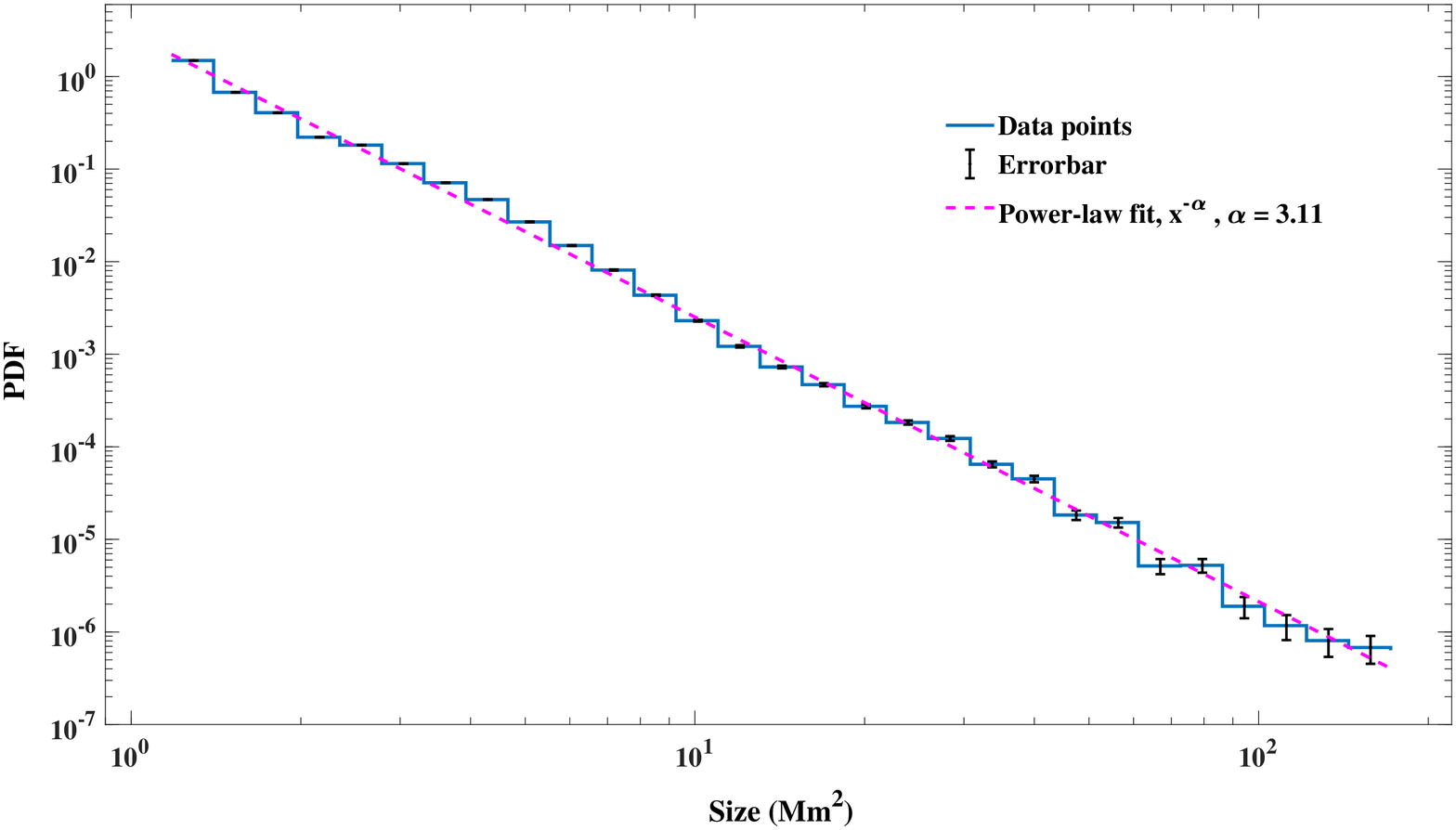}}
\centerline{\includegraphics[width=1\textwidth,clip=]{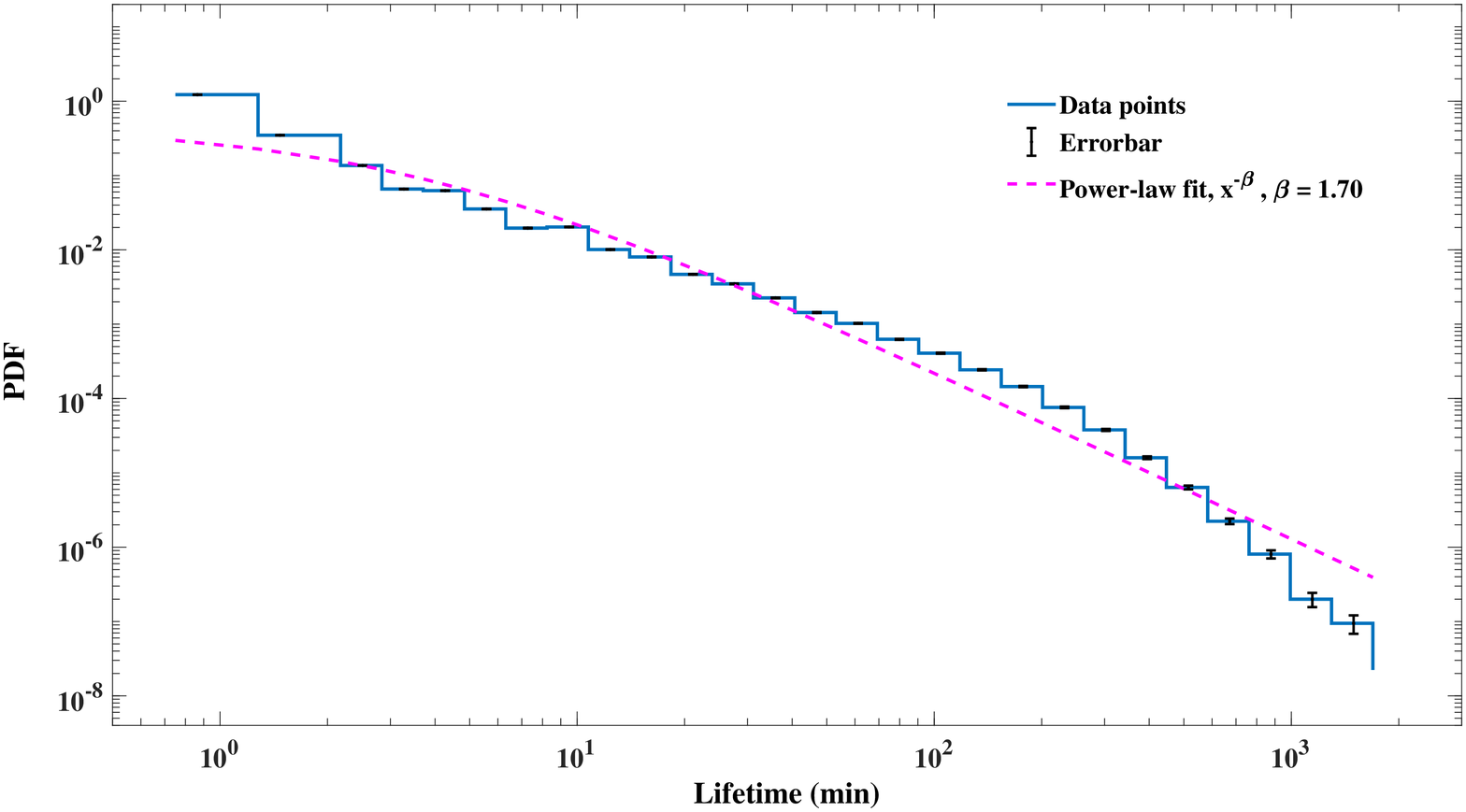}}
\caption{Distributions of the mean flux and the mean size of the magnetic elements during their lifetimes in the flaring AR are displayed on a log-log scale in the upper and middle panels, respectively. Distribution of lifetimes of magnetic elements in the flaring AR are plotted on a log-log scale in the lower panel. The power-law function was fitted to the distributions (dashed line). As it is shown in panels, the power-law exponents $\gamma$, $\alpha$, and $\beta$ were obtained to be about 2.36, 3.11, and 1.70, respectively.}
\end{figure}
\begin{figure}
\centerline{\includegraphics[width=1\textwidth,clip=]{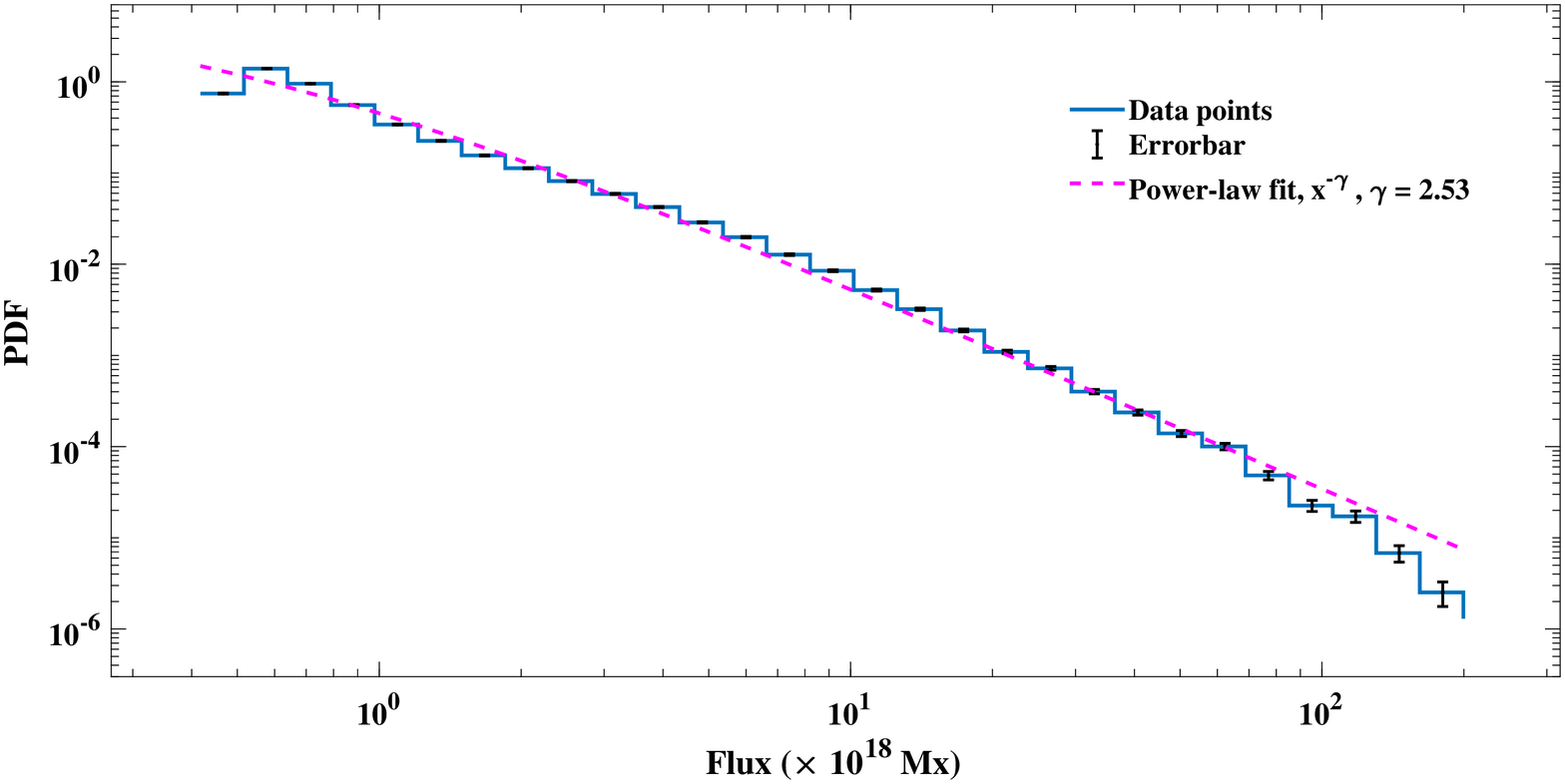}}
\centerline{\includegraphics[width=1\textwidth,clip=]{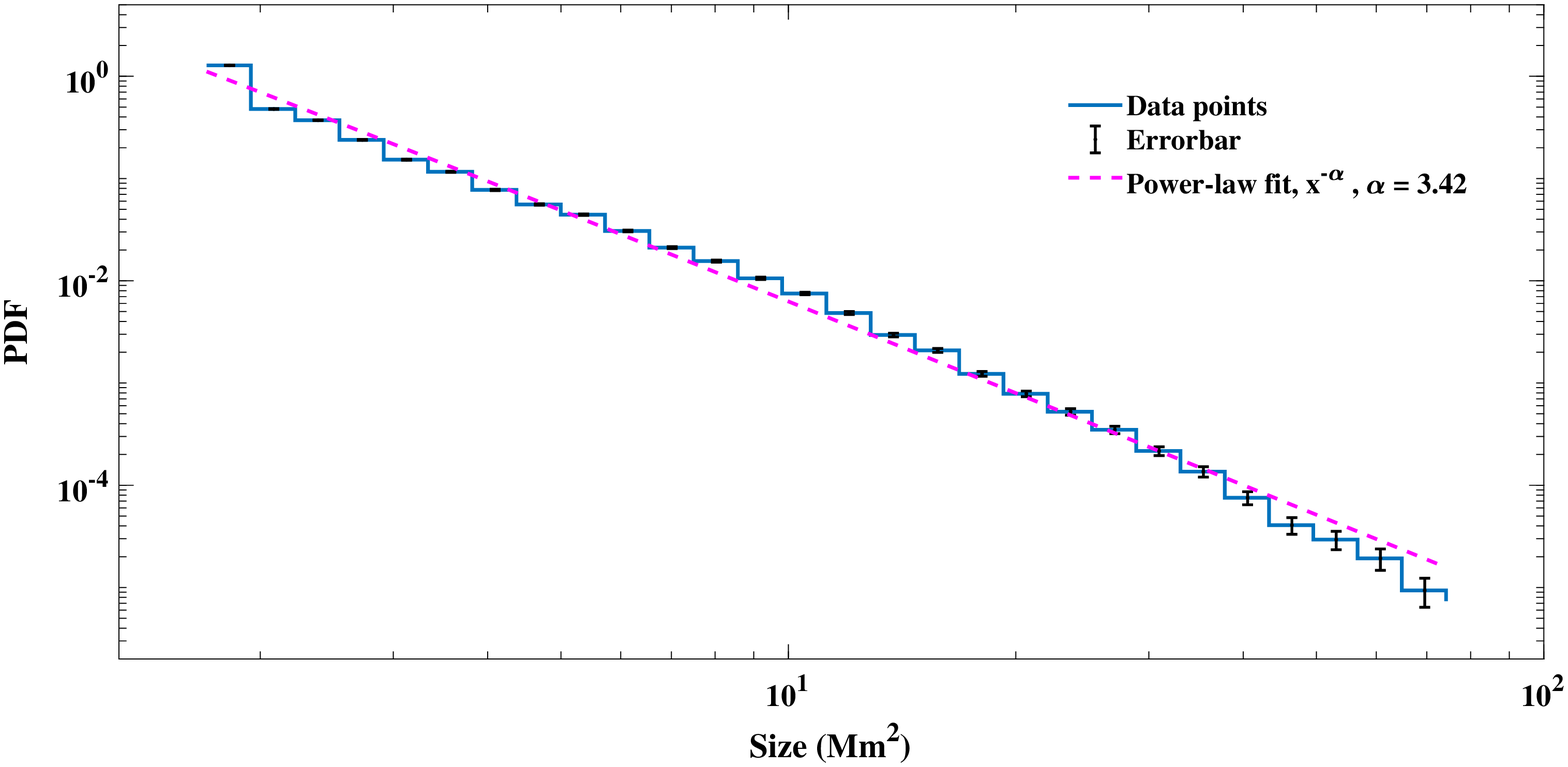}}
\centerline{\includegraphics[width=1\textwidth,clip=]{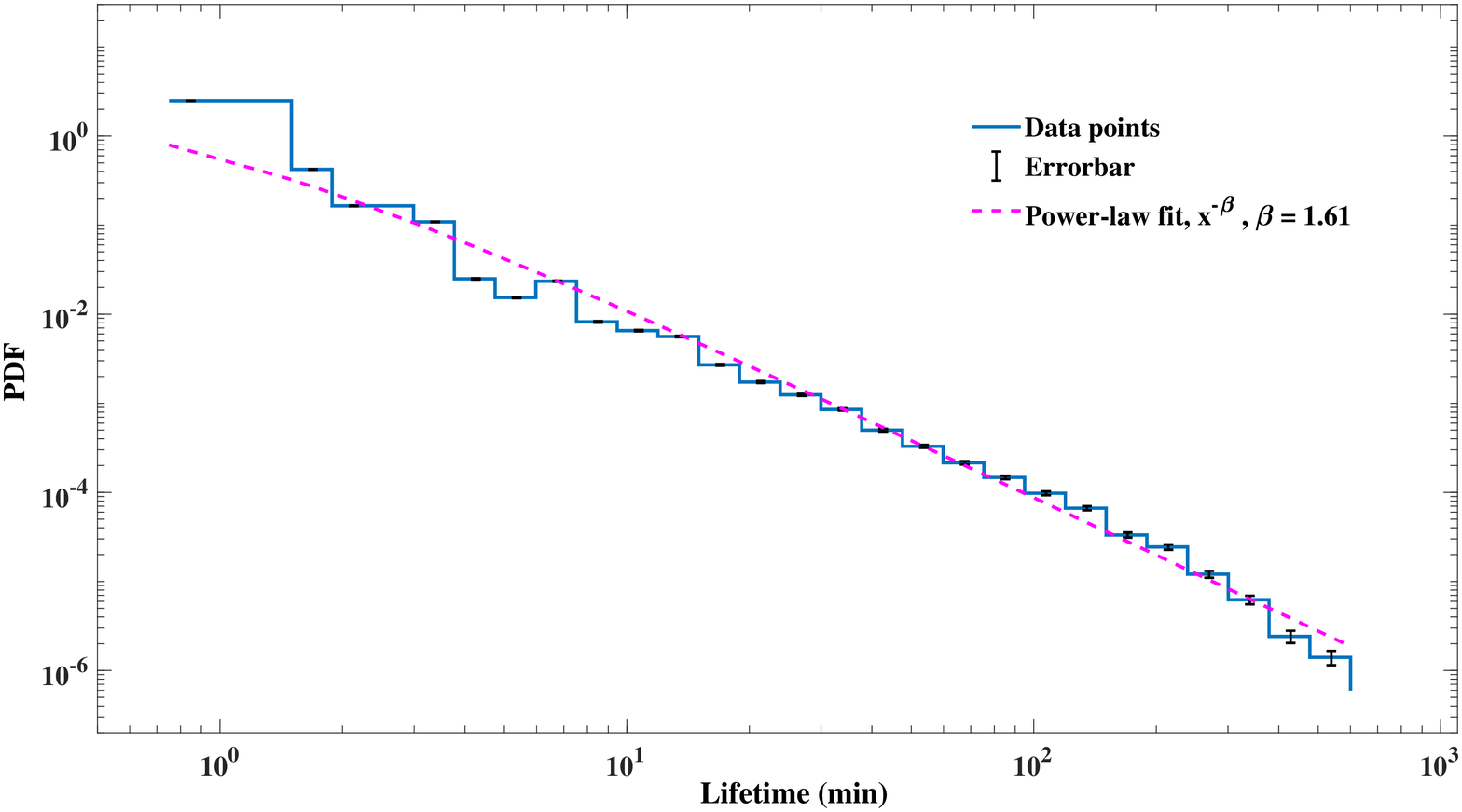}}
\caption{Distributions of the mean flux and the mean size of the magnetic elements during their lifetimes in the non-flaring AR are displayed on a log-log scale in the upper and the middle panels, respectively. lifetime Distribution of the magnetic elements in the non-flaring AR are plotted on a log-log scale in the lower panel. The power-law function was fitted to the distributions (dashed line). As it is shown in the panels, the power-law exponents $\gamma$, $\alpha$, and $\beta$ were obtained to be about 2.53, 3.42, and 1.61, respectively.}
\end{figure}
\subsection{Magnetic elements with maximum flux, size, and lifetime}
In Figure 7, patches with the maximum values of flux, size, and lifetime identified by the code in the flaring AR are presented. The maximum value of the mean flux belonged to a patch is $23.54\times10^{20}$ Mx (Figure 7, two first rows). This patch has a lifetime of about 165 minutes and a mean size of 296.12 Mm$^2$. The mean size's maximum value belonged to a patch is 299.65 Mm$^2$ (Figure 7, two middle rows). This patch has a lifetime of about 321 minutes and a mean flux of $23.47\times10^{20}$ Mx. Also, the lifetime's maximum value belonged to a patch is 2208 minutes (Figure 7, two last rows). This patch has a mean flux of $31.50\times10^{18}$ Mx and a mean size of 9.19 Mm$^2$. In Figure 8, patches with the maximum values of flux, size, and lifetime identified by the code in non-flaring AR are displayed. In this area, the mean flux's maximum value belonged to a patch is $2.46\times10^{20}$ Mx (Figure 8, two first rows). This patch has a lifetime of about 16 minutes and a mean size of 79.39 Mm$^2$. The mean size's maximum value belonged to a patch is 97.23 Mm$^2$ (Figure 8, two middle rows). This patch has a lifetime of about one hour and a mean flux of $2.02\times10^{20}$ Mx. Also, the lifetime's maximum value belonged to a patch is 756 minutes (Figure 8, two last rows). This patch has a mean flux of $6.77\times10^{18}$ Mx and  mean size of 9.73 Mm$^2$. The positively correlated behavior of flux and the size of magnetic elements is seen in both ARs. Furthermore, there are differences between the parameters obtained from the flaring AR and the non-flaring AR. The values of the maximum flux and size in the flaring AR are more remarkable than that of the non-flaring AR.
\\
\subsection{Empirical scaling relations between flux, size, and lifetime of the magnetic elements}
For magnetic elements that appeared in the flaring AR, the relations between flux [$F_{_{FAR}}$], size [$S_{_{FAR}}$], and lifetime [$T_{_{FAR}}$] are presented in Figure 9 (\textit{FAR} is the abbreviation of "flaring active region"). The scatter plot of the mean size [Mm$^{2}$] \textit{versus} mean flux [$10^{18}$ Mx] of patches over their lifetimes is shown on a log-log scale in Figure 9 (upper panel). It is seen that the greater fluxes are more scattered in size and more than 0.99\% of patches have fluxes lower than $10^{20}$ Mx. Also, more than 0.97\% of patches have sizes lower than 10 Mm${^2}$. Applying the linear fit to the mean values (red points) of each bin (0 -- 0.2, 0.2 -- 0.4 [$\times10^{18}$ Mx], etc.) revealed that the relation between the size and flux could be expressed as $S_{_{FAR}} \propto F_{_{FAR}}^{a}$, with a best-fit parameter of $a = 0.66\pm0.01$ ($R$-square = 0.91). A closer look exposes that there may be a break in this scatter plot. Fitting a broken double linear function shows a gentle slope of about 0.43 (Figure 9, yellow line in upper panel) between the flux and size, for fluxes $< 8\times10^{19}$ Mx and sizes $< 15$ Mm$^2$. For greater fluxes and sizes, the slope takes value of about 0.93 (Figure 9, upper panel). It may suggest that after this limit, the correlated behavior of the flux and size significantly increases in the flaring AR. The scatter plot of the mean flux \textit{versus} the patches' lifetime is presented on a log-log scale in Figure 9 (middle panel). We can see that patches at any lifetime range, especially in shorter intervals, are more scattered in the flux. We applied the linear fit to the mean values (red points) of each bin (0 -- 1, 1 -- 2 [minutes], etc.) to find the relation between the mean flux [$10^{18}$ Mx] of patches during their lifetimes and the lifetimes [minutes]. The relation is $F_{_{FAR}} \propto T_{_{FAR}}^{c}$, wherein $c = 0.48\pm0.04$ ($R$-square = 0.38). The scatter plot of the mean size \textit{versus}  the patches' lifetime is presented on a log-log scale in Figure 9 (lower panel). As same as the last panel, we see that patches with various lifetimes have a large scatter in their sizes. The linear fit to the mean values (red points) of each bin (0 -- 1, 1 -- 2 [minutes], etc.) was applied to extract the power-law relation as $S_{_{FAR}} \propto T_{_{FAR}}^{e}$ where the parameter $e$ equals to $0.32\pm0.02$ ($R$-square = 0.50).

In the same manner as Figure 9, for magnetic elements appeared in the non-flaring AR, the relations between flux [$F_{_{AR}}$], size [$S_{_{AR}}$], and lifetime [$T_{_{AR}}$] are presented in Figure 10. The scatter plot of the mean size [Mm$^{2}$] \textit{versus} mean flux [$10^{18}$ Mx] of patches over their lifetimes is shown on a log-log scale in Figure 10 (upper panel). It is seen that the greater fluxes are more scattered in size and more than 0.99\% of patches have fluxes lower than $10^{20}$ Mx. Also, more than 0.96\% of patches have sizes lower than 10 Mm${^2}$. Applying the linear fit to the mean values (red points) of each bin (0 -- 0.2, 0.2 -- 0.4 [$\times10^{18}$ Mx], etc.) revealed that the relation between the size and flux could be expressed as $S_{_{AR}} \propto F_{_{AR}}^{g}$, with a best-fit parameter of $g = 0.64\pm0.02$ ($R$-square = 0.90). As same as Figure 9, fitting a broken double linear function reveals that for fluxes $< 2\times10^{19}$ Mx and sizes $< 15$ Mm$^2$, the slope is obtained to be about 0.55 (Figure 10, yellow line in upper panel). For the greater fluxes and sizes, the slope takes the value of about 0.81 (Figure 10, green line in upper panel). The scatter plot of the mean flux \textit{versus} the patches' lifetime is presented on a log-log scale in Figure 10 (middle panel). We can see that patches at any ranges of lifetimes are more scattered in flux. We applied the linear fit to the mean values (red points) of each bin (0 -- 1, 1 -- 2 [minutes], etc.) to find the relation between the mean flux [$10^{18}$ Mx] of patches during their lifetimes and the lifetimes [minutes]. The relationship is $F_{_{AR}} \propto T_{_{AR}}^{i}$, wherein $i = 0.37\pm0.06$ ($R$-square = 0.16). The scatter plot of the mean size \textit{versus} the patches' lifetime is presented on a log-log scale in Figure 10 (lower panel). Similar to the previous panel, we see that patches with various lifetimes have a large scatter in their sizes. The linear fit to the mean values (red points) of each bin (0 -- 1, 1 -- 2 [minutes], etc.) was applied to extract the power-law relation as $S_{_{AR}} \propto T_{_{AR}}^{k}$ where the parameter $k$ equals to $0.23\pm0.03$ ($R$-square = 0.24).

Let us assume two power-law distributions with functionality of variables $x$ and $y$ as the forms of $N(x) \propto x^{-n_x}$ and $N(y) \propto y^{-n_y}$, respectively. The relation between two variables can be specified by $y \propto x^m$, wherein $m$ is as follows (Aschwanden 2011)
\begin{equation}\label{exponentm}
m = \frac{n_x-1}{n_y-1}.
\end{equation}
In the case of linear proportionality of variables $x$ and $y$, the parameter $m$ approaches unity. Since we obtained the exponent $m$ by applying a linear fit to the logarithmic scatter plots shown in Figures 9 and 10, the consistency between the resultant power-law exponents can be checked by the Eq. 5.

For the flaring AR, the power-law indices of flux ($n_F$), size ($n_S$), and lifetime ($n_T$) extracted from frequency distributions (see Figure 5) were obtained  to be about 2.36, 3.11, and 1.70, respectively. Replacing a pair of the power indices in Eq. 5, the corresponding scaling laws are achieved
\begin{equation}\label{scalawar}
S_{FAR} \propto F_{FAR}^{0.64},~~~~~~ F_{FAR} \propto T_{FAR}^{0.51},~~~~~~ S_{FAR} \propto T_{FAR}^{0.33}.
\end{equation}
For the non-flaring AR, these power-law indices which were extracted from frequency distributions (see Figure 6) were about 2.53, 3.42, and 1.61, respectively.
\begin{equation}\label{scalawnar}
S_{AR} \propto F_{AR}^{0.63},~~~~~~ F_{AR} \propto T_{AR}^{0.39},~~~~~~ S_{AR} \propto T_{AR}^{0.25}.
\end{equation}
As it is seen, these obtained scaling values for the flaring AR and the non-flaring-AR are in good agreement with results extracted from Figures 9 and 10, respectively.

\subsection{Empirical scaling relation between size and growth rate of magnetic elements}
As it was mentioned in Section 1, (Otsuji \textit{et al.} 2011) obtained a power-law relation between the maximum flux of positive magnetic polarities $F_{max}$ and their flux growth rate $<\frac{dF}{dt}>$ as follows
\begin{equation}\label{scalawgrowthf}
<\frac{dF}{dt}> \propto F_{max}^{0.5},
\end{equation}
where $t$ is the time of flux growth and $\left\langle \right\rangle$ denotes the mean value. In Figure 11, the scatter plot of the mean size of the positive polarities \textit{versus} their maximum flux on a log-log scale for the flaring AR (upper panel) and for the non-flaring AR (lower panel) is plotted. As it is seen, the scaling relation between the mean size of the positive polarities and their maximum flux takes the relation as $S \propto F_{max}^{0.64}$ ($R$-square = 0.90) for the flaring AR and $S \propto F_{max}^{0.68}$ ($R$-square = 0.93) for the non-flaring AR. The relation between the flux growth rate of the positive polarities and their size can be yielded with the use of Eq. 8 as follows
\begin{eqnarray}\label{scalawgrowths}
<\frac{dF}{dt}> \propto S^{0.78}~~~~ \rm{for~the~flaring~AR},~~~~\nonumber\\
<\frac{dF}{dt}> \propto S^{0.73}~~~~ \rm{for~the~non-flaring~AR}.
\end{eqnarray}
This power-law relation indicates that there is strong correlation between the maximum flux growth rate of the positive polarities and their size in both ARs.

\begin{figure}
\centerline{\includegraphics[width=1.02\textwidth,clip=]{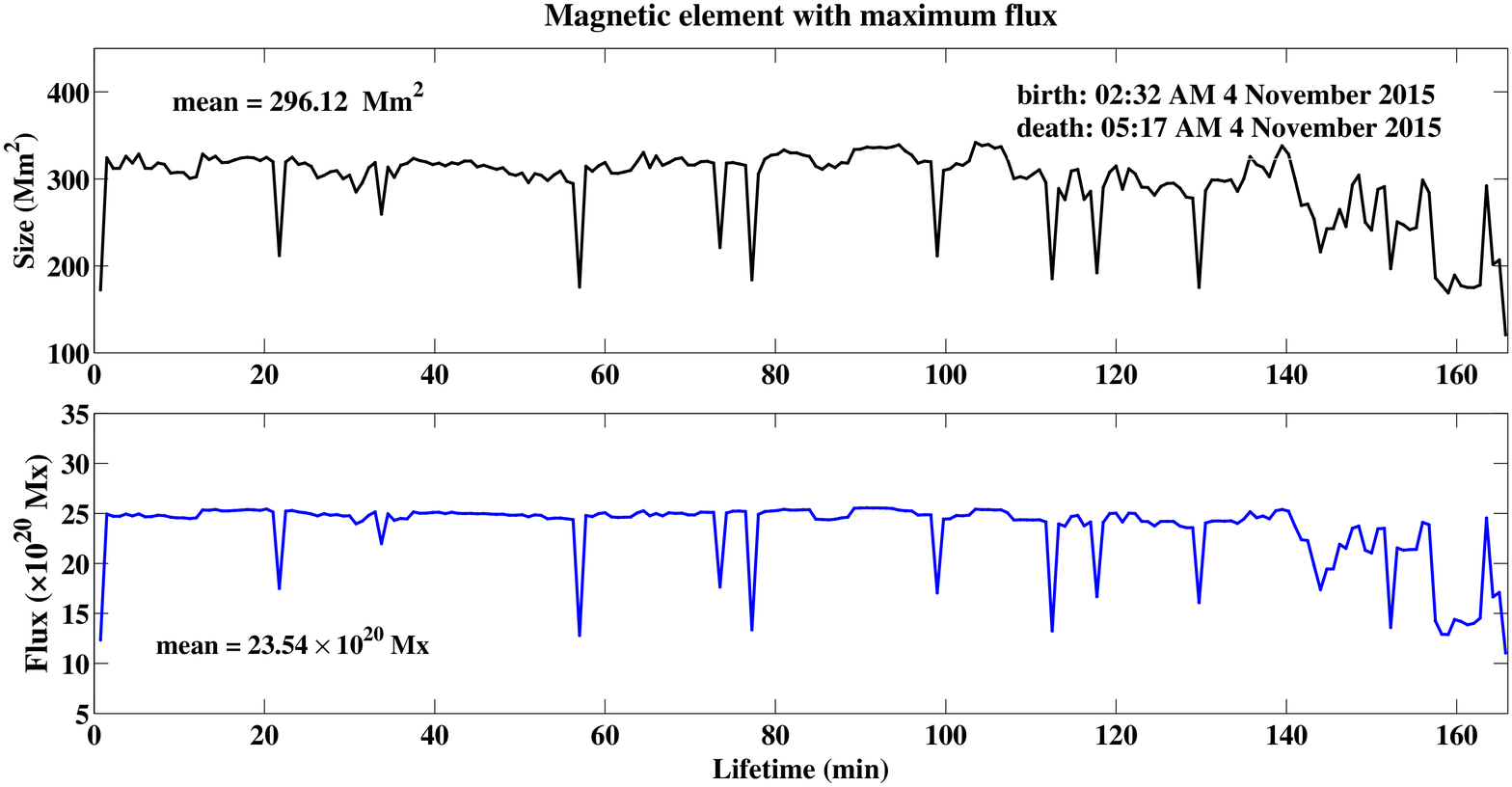}}
\centerline{\includegraphics[width=1.02\textwidth,clip=]{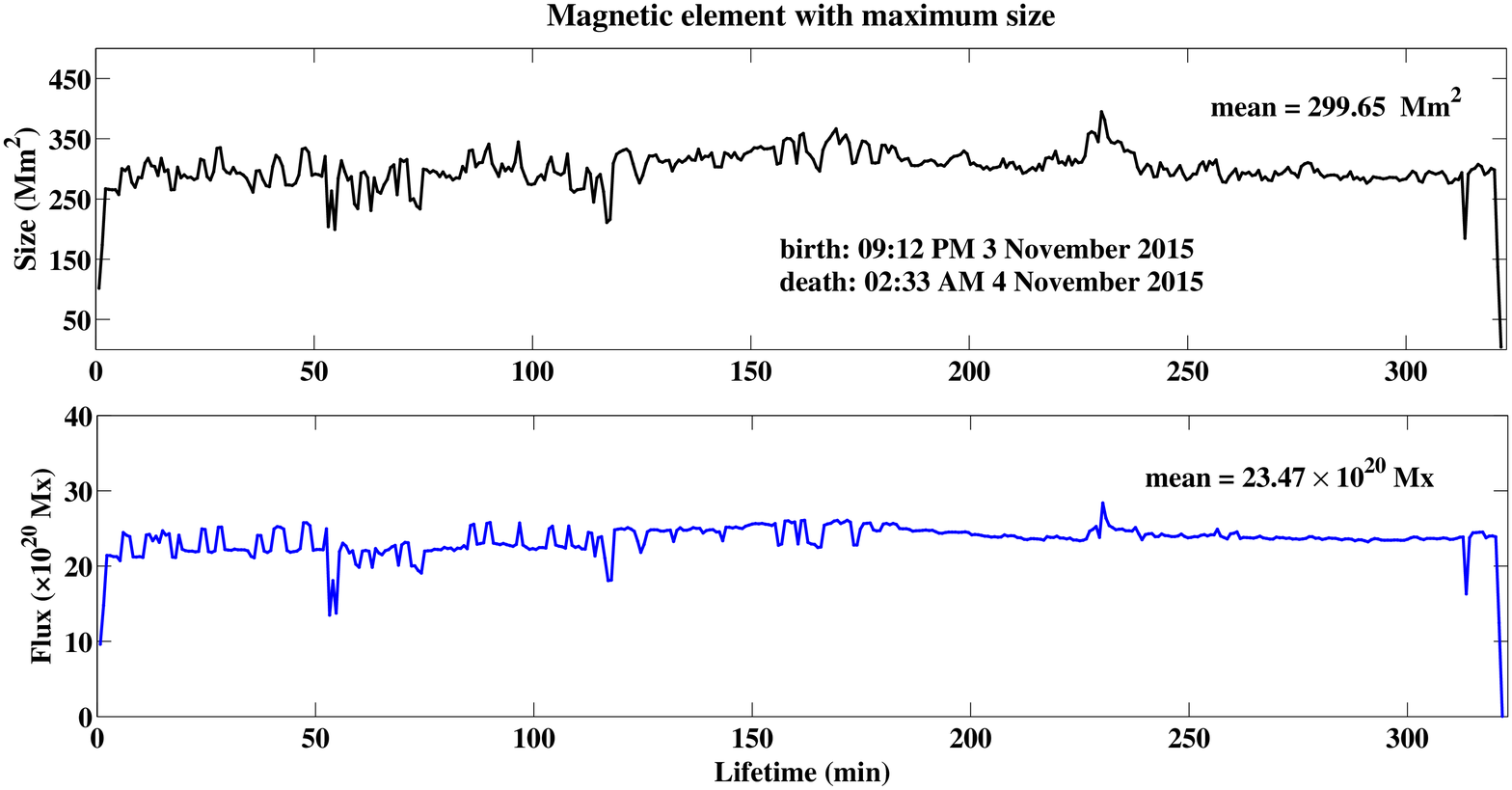}}
\centerline{\includegraphics[width=1.02\textwidth,clip=]{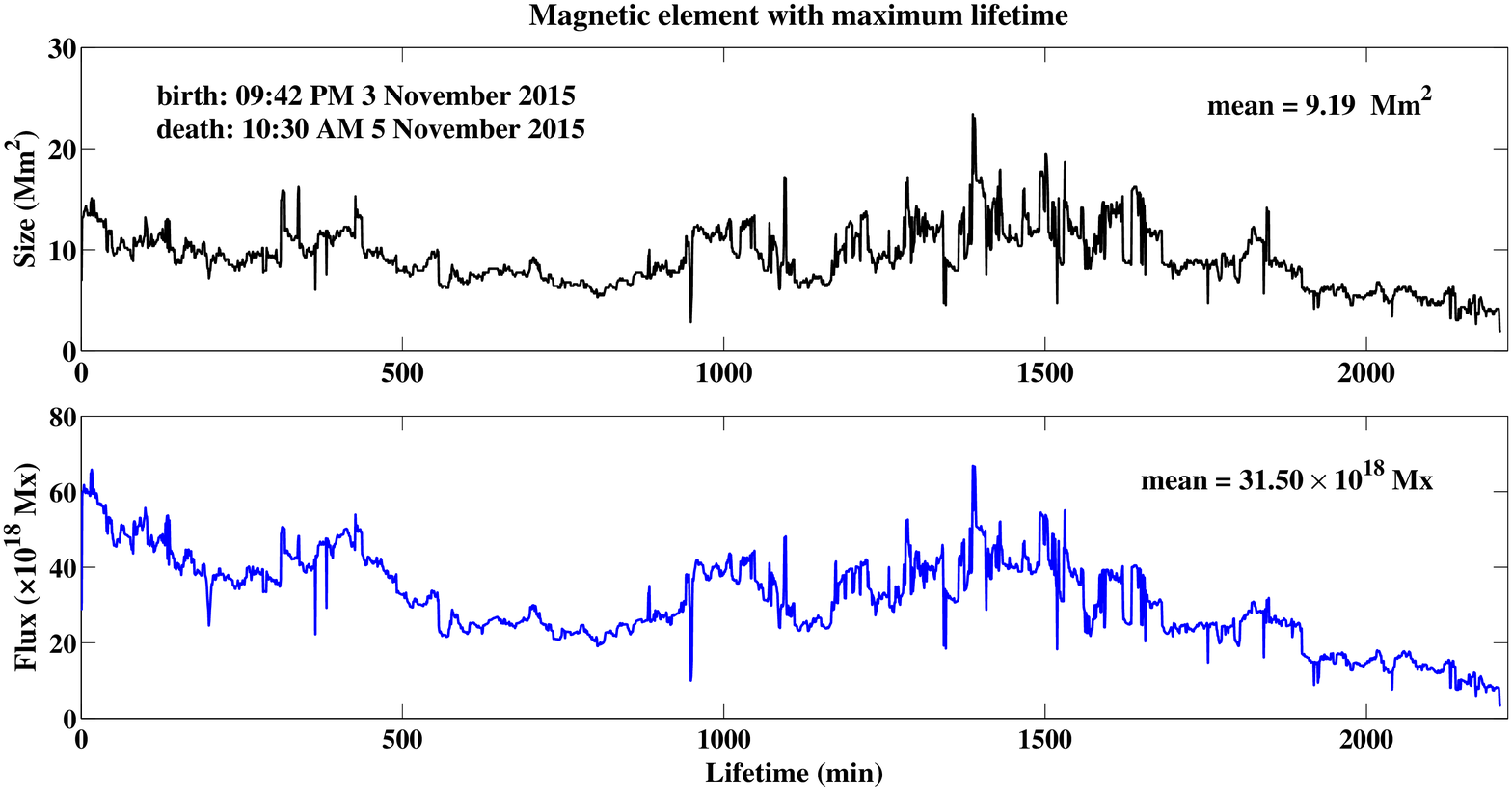}}
\caption{Size variations of a patch with the highest flux ($23.54\times10^{20}$ Mx) in the flaring AR and its flux variations in time are presented in two first panels, respectively. Size variations of a patch with the largest area (299.65 Mm$^2$) in the flaring AR and its flux variations in time are presented in two middle panels, respectively. Size variations of the most long-lived patch (2208 minutes) in the flaring AR and its flux variations in time are presented in two last panels.}
\end{figure}
\begin{figure}
\centerline{\includegraphics[width=1.02\textwidth,clip=]{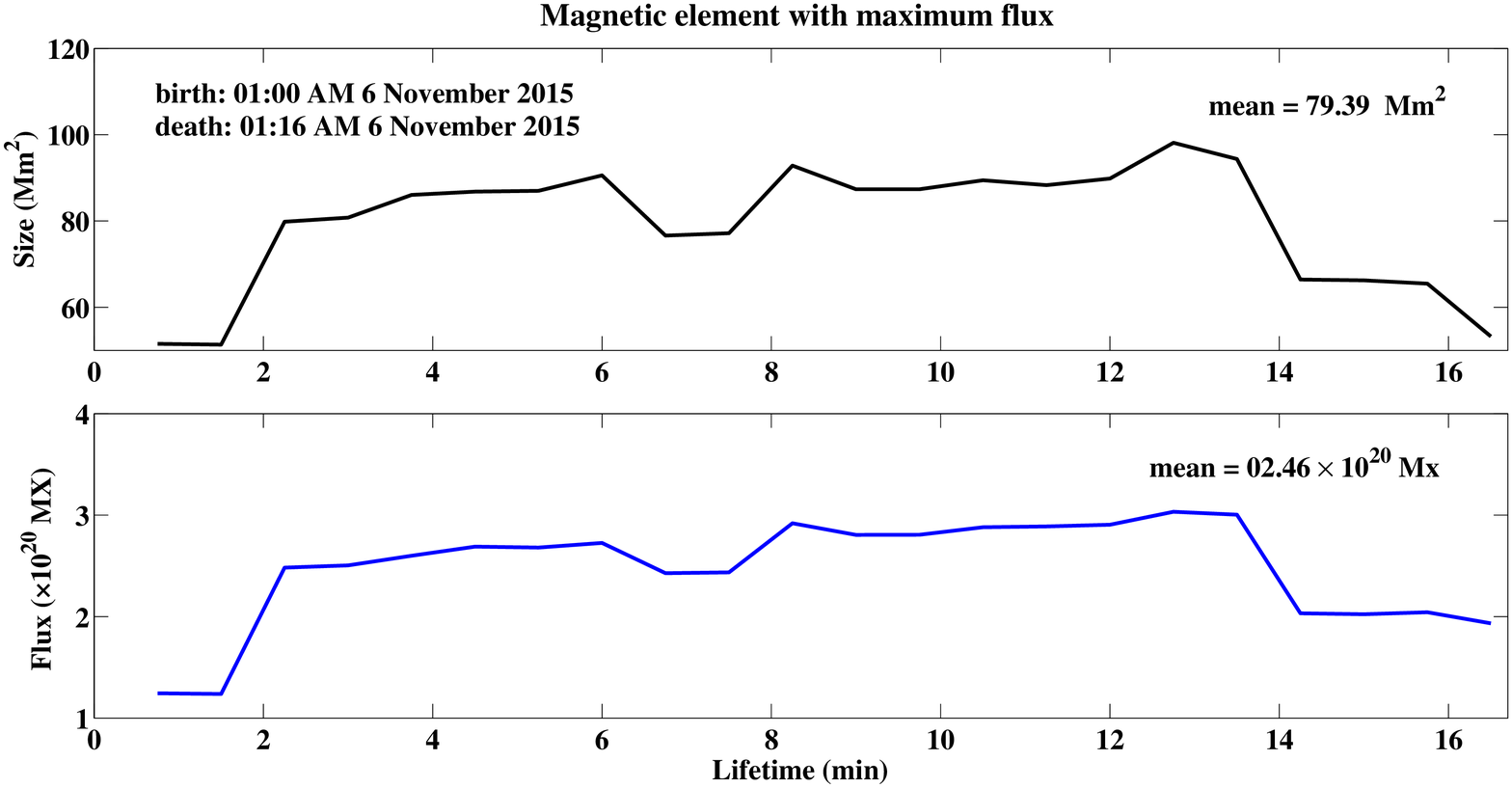}}
\centerline{\includegraphics[width=1.02\textwidth,clip=]{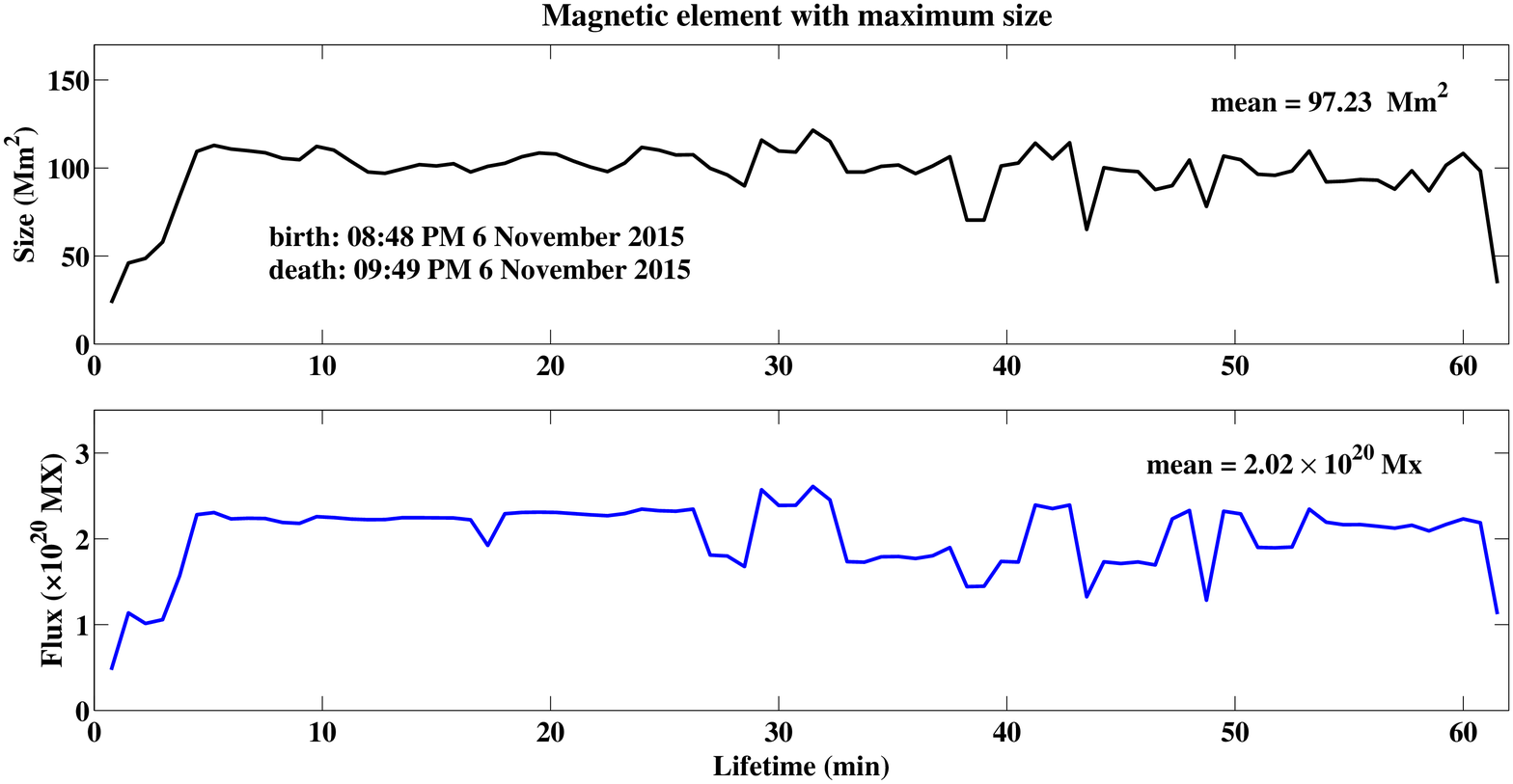}}
\centerline{\includegraphics[width=1.02\textwidth,clip=]{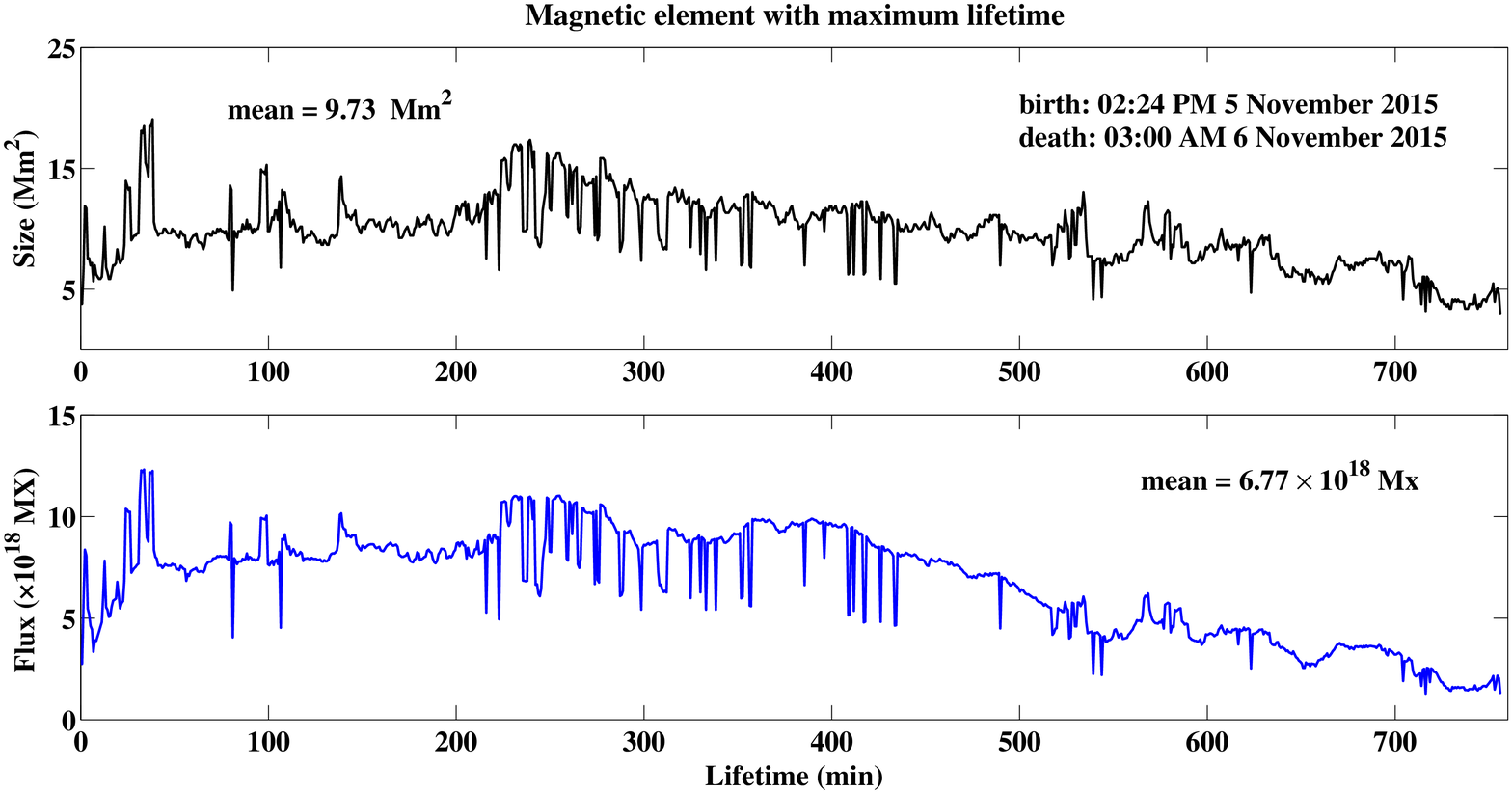}}
\caption{Size variations of a patch with the highest flux ($2.46\times10^{20}$ Mx) in the non-flaring AR and its flux variations in time are presented in two first panels, respectively. Size variations of a patch with the largest area (97.23 Mm$^2$) in the non-flaring AR and its flux variations in time are presented in two middle panels, respectively. Size variations of the most long-lived patch (12 hours and 36 minutes) in the non-flaring AR and its flux variations in time are presented in two last panels.}
\end{figure}
\begin{figure}
\centerline{\includegraphics[width=0.9\textwidth,clip=]{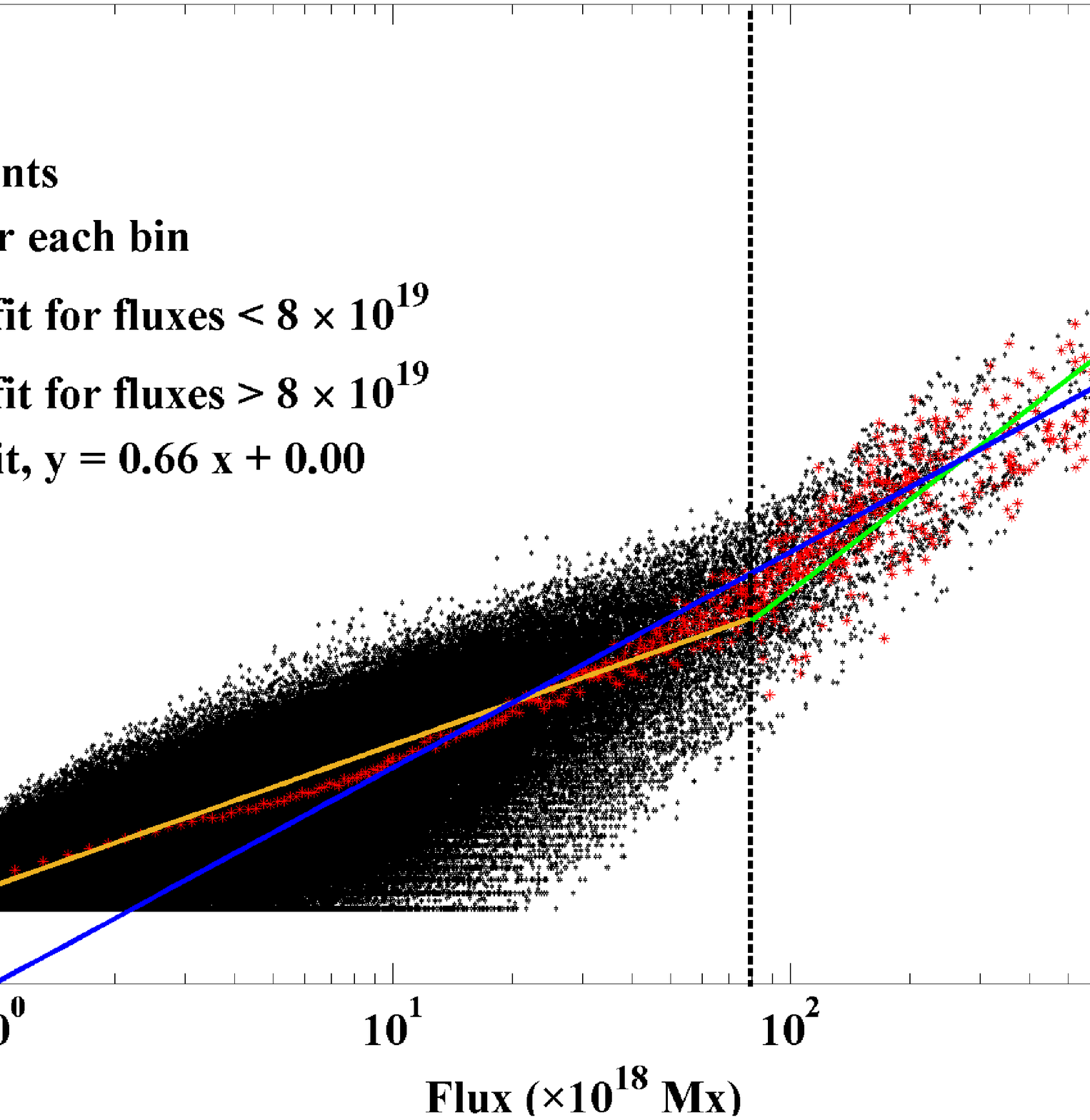}}
\centerline{\includegraphics[width=0.98\textwidth,clip=]{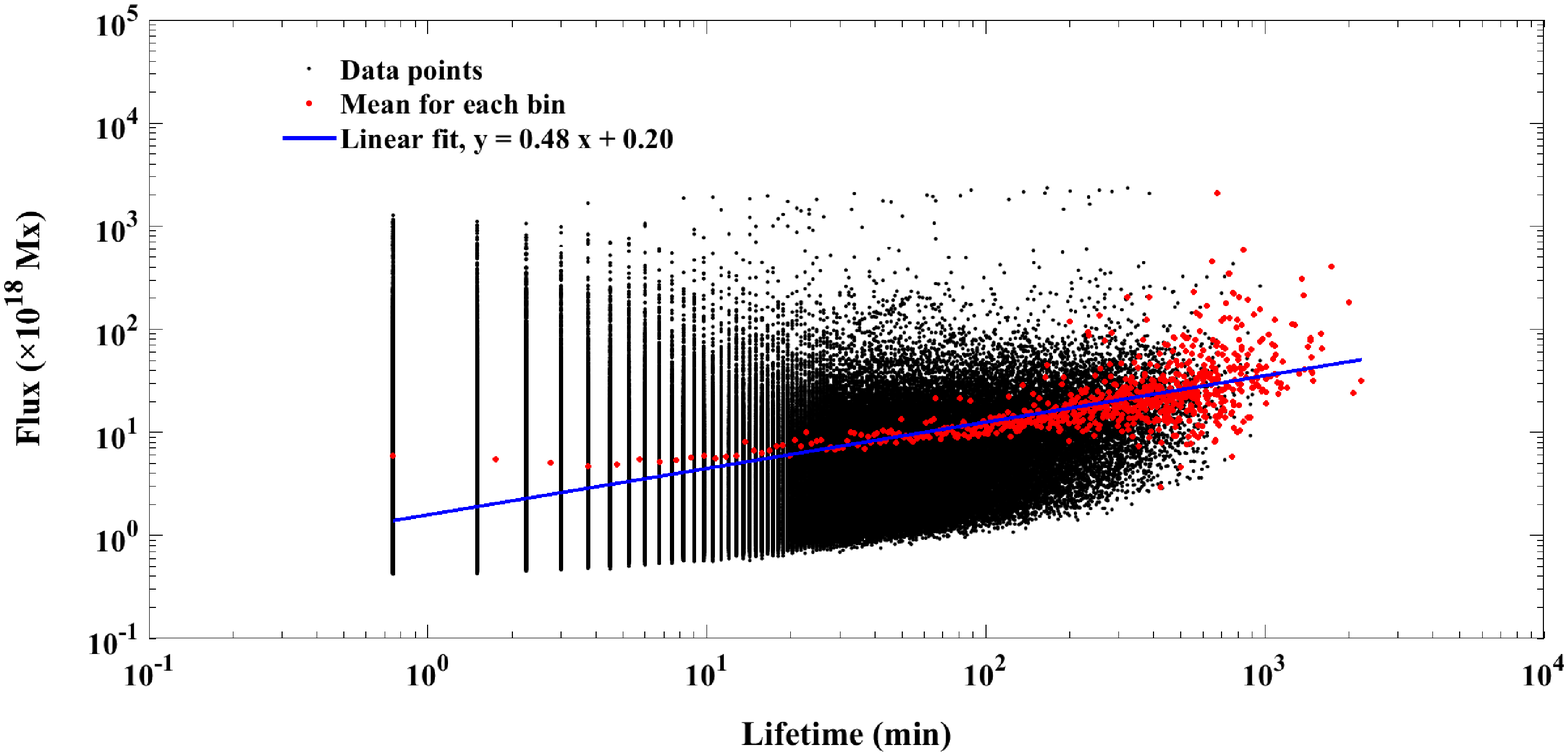}}
\centerline{\includegraphics[width=0.98\textwidth,clip=]{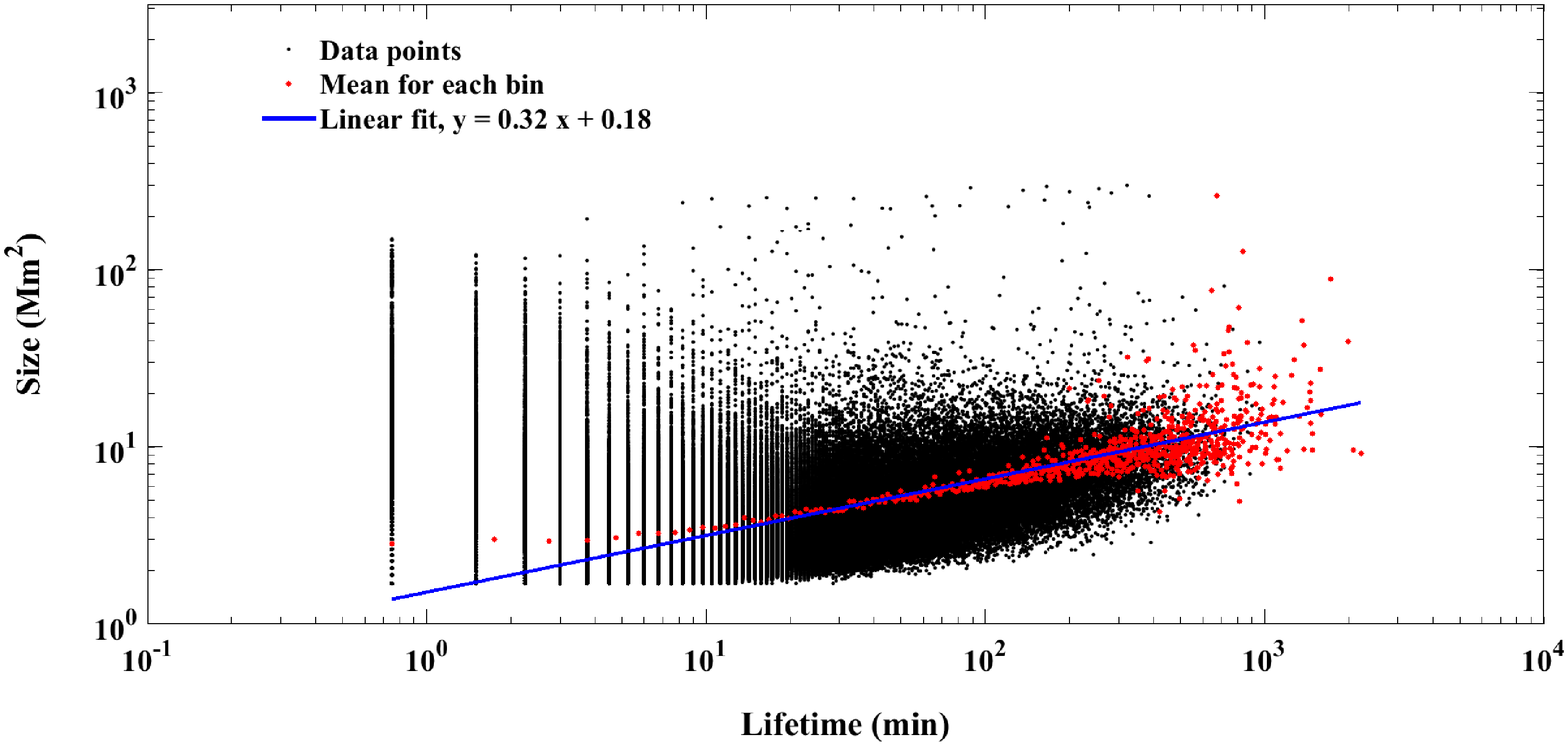}}
\caption{Relation between the mean flux of the patches [Mx] and their mean size [Mm$^2$] during their lifetimes for the whole range of the fluxes (upper panel), relation between the mean flux of the patches [Mx] and their lifetime [minutes] (middle panel), and relation between the mean size of the patches [Mm$^2$] and their lifetime [minutes] (lower panel) on a log-log scale as they appeared in the flaring AR. The mean values of each bin in the upper panel (0 -- 0.2, 0.2 -- 0.4 [$\times 10^{18}$ Mx], etc.) and in two later panels (0 -- 1, 1 -- 2 [minutes], etc.) for patches are shown by red markers. The blue-solid lines display the linear fits as $\log(S) = a\log(F) + b$ with $a = 0.66 \pm 0.01$ and $b = 0 \pm 0.02$ (upper panel), $\log(F) = c\log(T) + d$ with $c = 0.48 \pm 0.04$ and $d = 0.20 \pm$ 0.01 (middle panel), and $\log(S) = e\log(T) + f$ with $e = 0.32 \pm 0.02$ and $f = 0.18 \pm 0.05$ (lower panel). A closer look to the upper panel illustrates a broken double linear function. For fluxes smaller and greater than $8\times10^{19}$ Mx, the slopes were obtained as 0.43 and 0.93, respectively.}
\end{figure}
\begin{figure}
\centerline{\includegraphics[width=0.9\textwidth,clip=]{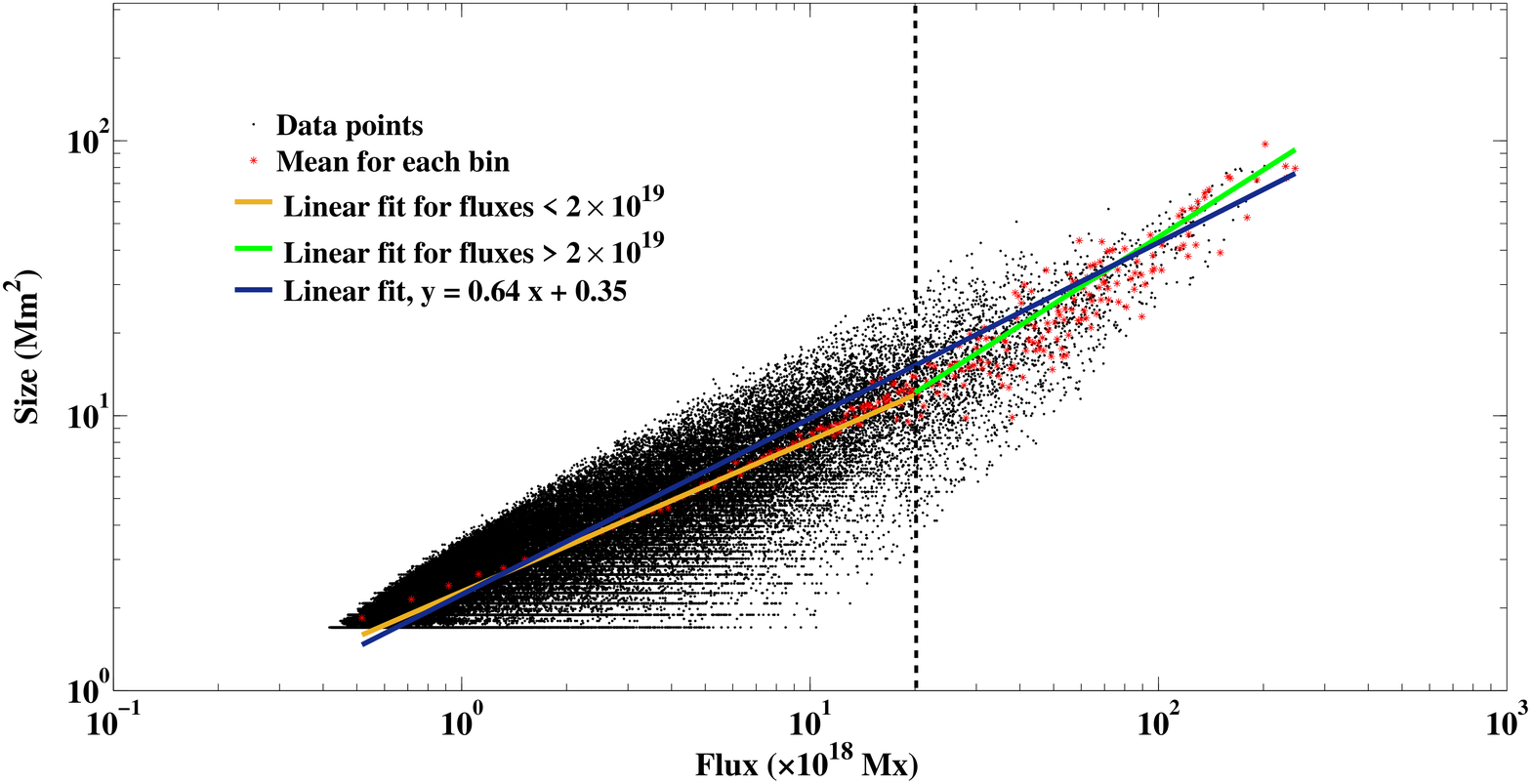}}
\centerline{\includegraphics[width=0.98\textwidth,clip=]{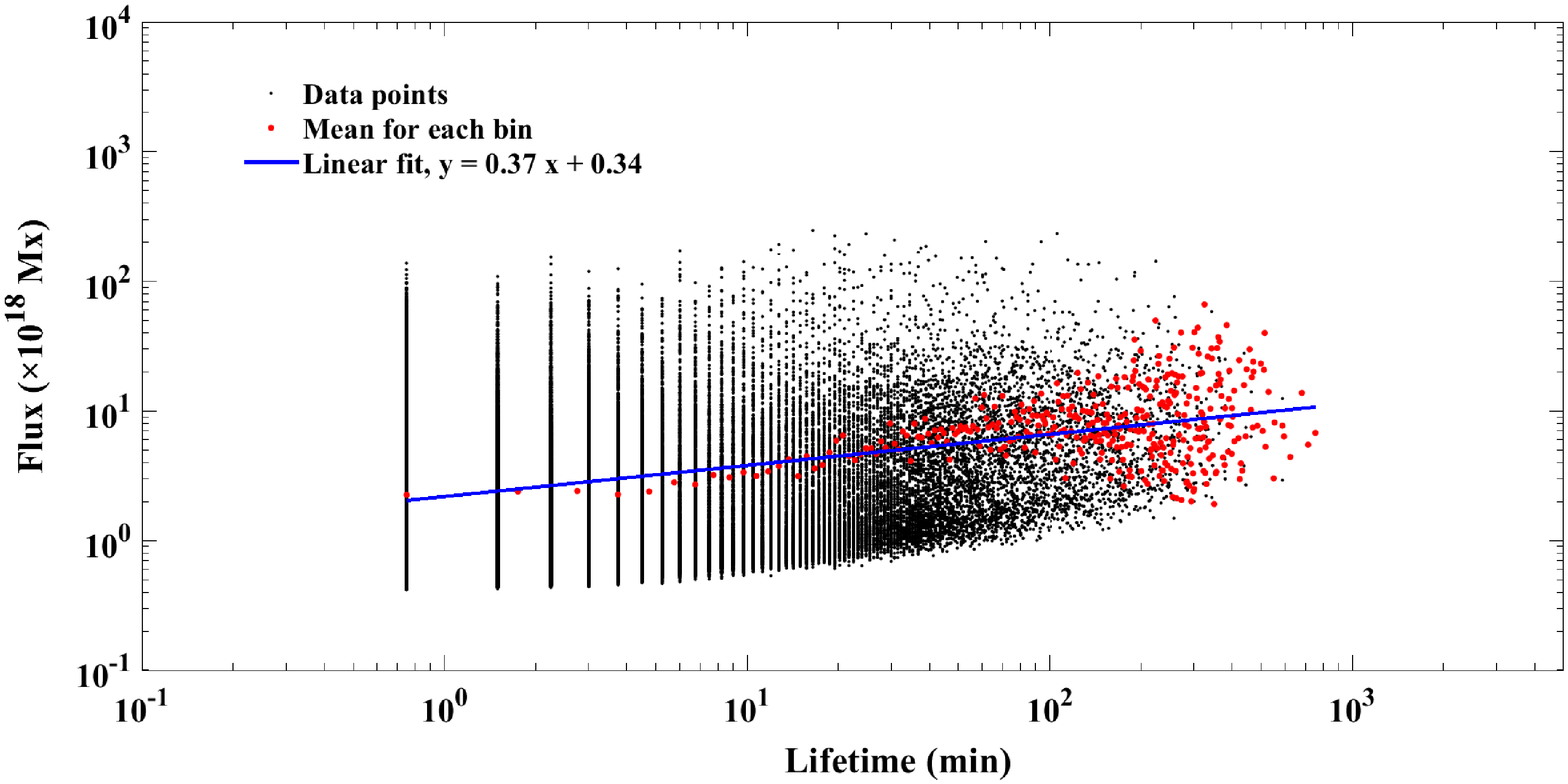}}
\centerline{\includegraphics[width=0.98\textwidth,clip=]{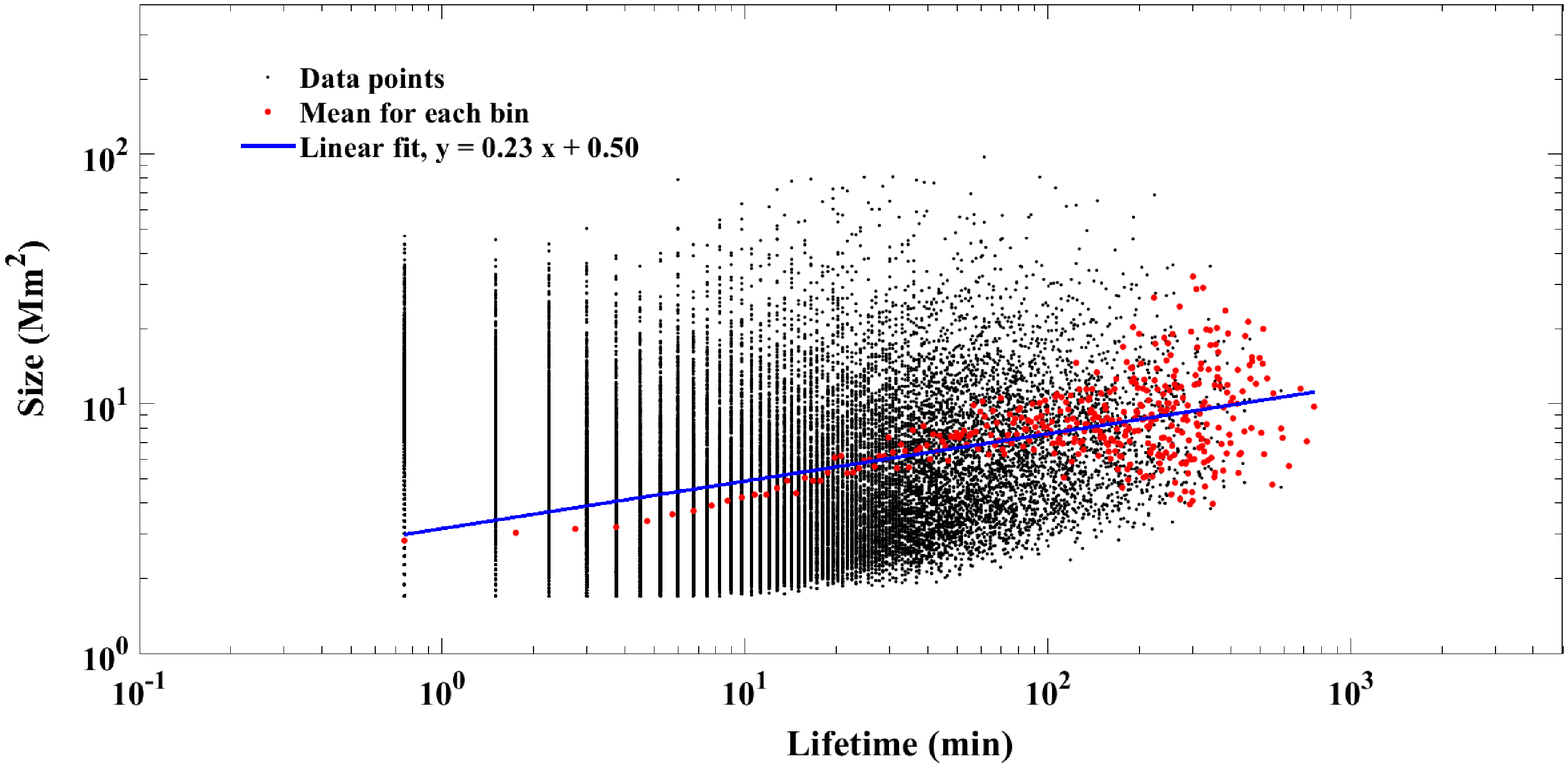}}
\caption{Relation between the mean flux of the patches [Mx] and their mean size [Mm$^2$] during their lifetimes for the whole range of the fluxes (upper panel), relation between the mean flux of the patches [Mx] and their lifetime [minutes] (middle panel), and relation between the mean size of patches [Mm$^2$] and their lifetime [minutes] (lower panel) on a log-log scale as they appeared in the non-flaring AR. The mean values of each bin in the upper panel (0 -- 0.2, 0.2 -- 0.4 [$\times 10^{18}$ Mx], etc.) and in two later panels (0 -- 1, 1 -- 2 [minutes], etc.) for patches are shown by red markers. The blue-solid lines show the linear fits as $\log(S) = g\log(F) + h$ with $g = 0.64 \pm 0.02$ and $h = 0.30 \pm 0.03$ (upper panel), $\log(F) = i\log(T) + j$ with $i = 0.37 \pm 0.06$ and $j = 0.34 \pm 0.15$ (middle panel), and $\log(S) = k\log(T) + l$ with $k = 0.23 \pm 0.03$ and $l = 0.50 \pm 0.08$ (lower panel). A closer look to the upper panel illustrates a broken double linear function. For fluxes smaller and greater than $2\times10^{19}$ Mx, the slopes were obtained as 0.55 and 0.81, respectively.}
\end{figure}
\begin{figure}
\centerline{\includegraphics[width=1.05\textwidth,clip=]{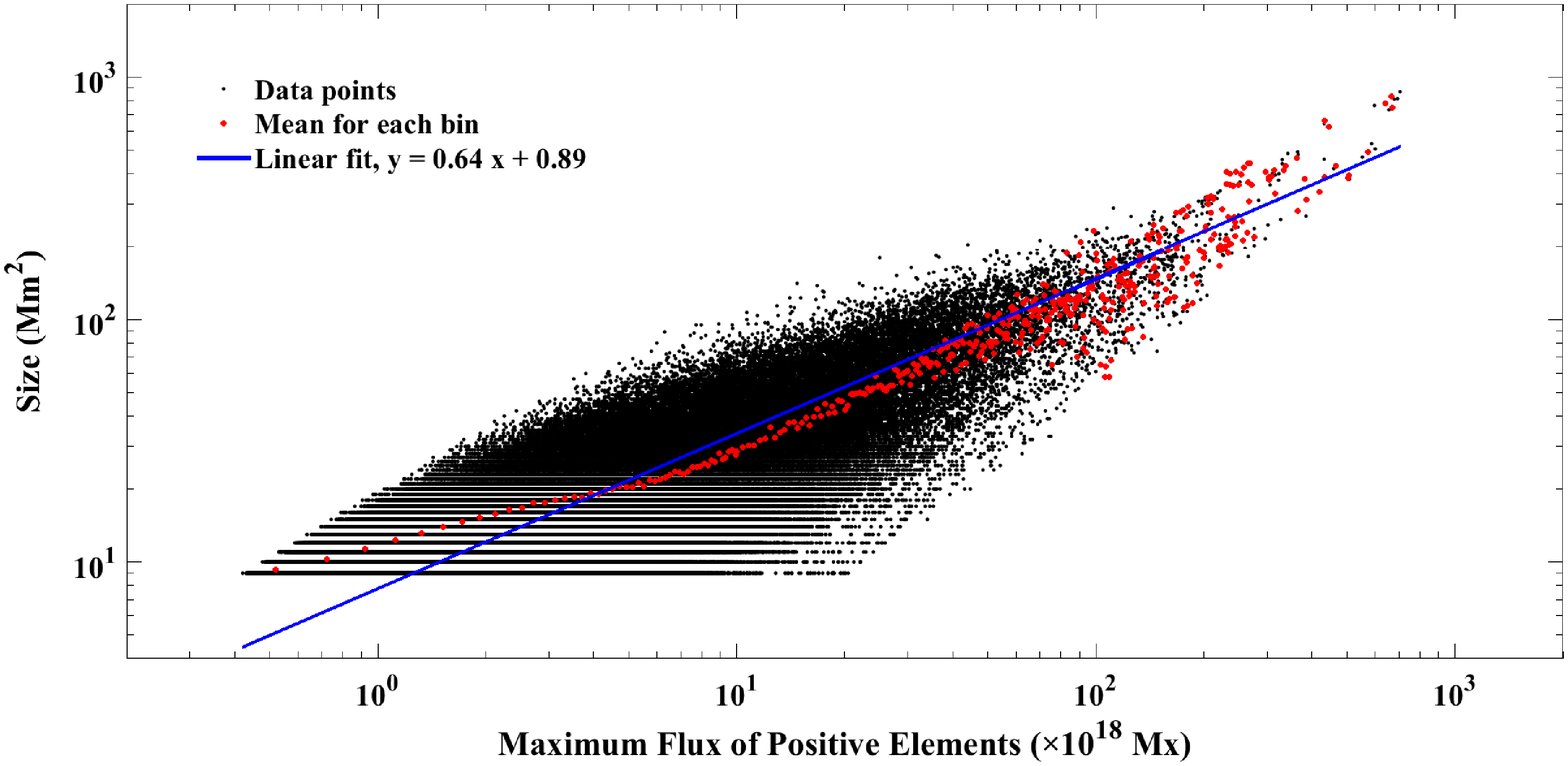}}
\centerline{\includegraphics[width=1.05\textwidth,clip=]{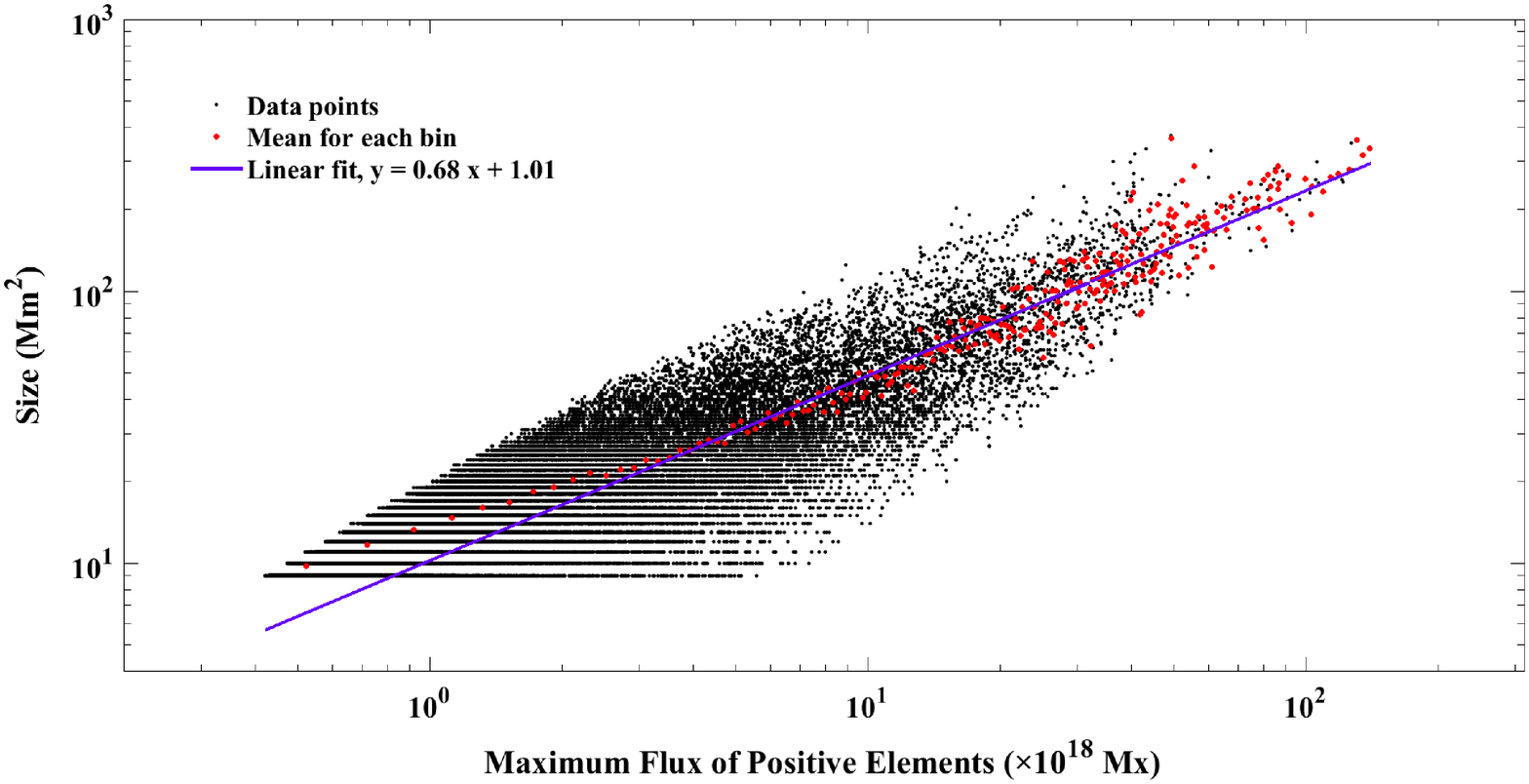}}
\caption{Relation between the maximum flux of the positive polarities [Mx] and their mean size [Mm$^2$] during their lifetimes in the flaring AR (upper panel) and in the non-flaring AR (lower panel). The mean values of each bin (0 -- 0.2, 0.2 -- 0.4 [$\times 10^{18}$ Mx], etc.) for polarities are shown by red points. The blue-solid lines show the linear fits as $\log(S) = p\log(F_{max}) + q$ with $p = 0.64 \pm 0.01$ and $q = 0.89 \pm 0.03$ for the flaring AR (upper panel), and with $p = 0.68 \pm 0.02$ and $q = 1.01 \pm 0.03$ for the non-flaring AR (lower panel).}
\end{figure}
\section{Conclusions}
The number of 347502 and 114063 photospheric magnetic elements was automatically detected in the flaring AR (NOAA 12443) and in the non-flaring AR (NOAA 12446), respectively, over the HMI consecutive sub-images. Among identified patches, the largest one's label belonged to a patch with a mean area of about 300 Mm$^2$. Furthermore, we found a patch with the maximum flux of $23.54\times10^{20}$ Mx. The most long-lived patch has a lifetime of 2208 minutes ($\sim$ 37 hours). Expectedly, all these patches were appeared in the flaring AR. Statistical analyzes show that both the mean value of the area filling factor and the flux summation of the magnetic patches in the flaring AR are more than those of the non-flaring AR. The greater value of the number of elements in the flaring AR implies the more complexity of the patches' system in this area, which may rise the possibility of flare occurrence. All frequency distributions of flux, size, and lifetime in both ARs follow power-law functions indicating their scale-invariant behavior. Data-truncation effects caused by applying detection thresholds on both size and magnetic field of the patches can lead to a little deviations from ideal power laws, as seen in the distributions. According to the fitted broken double linear functions to the scatter plots of flux and size, it can be inferred that the start flux range of significant increases in the correlated behavior of flux and size of patches is 2 -- 8 [$\times 10^{19}$ Mx].

As a comparison between the power-law exponents of the flux-distribution and size-distribution of the patches in both ARs, the greater magnitude value of slopes of both distributions in the non-flaring AR suggests that the contribution of small scale events in flux and size belonging to this area is more than that of the flaring AR.

On the contrary of the anti-correlated behavior which is observed in the filling factor of the negative and positive polarities (with the Pearson value of -0.88) in the flaring AR, the filling factor of the positive and negative polarities in the non-flaring AR are significantly correlated (with the Pearson correlation value of 0.97). The weak anti-correlated behavior between the total area filling factors and the total number of patches in both ARs would imply that the area filling factor decreases by increasing the number of patches. On the other hand, the Pearson correlation values between the time series of the positive and negative flux in the flaring AR ($\sim$ 0.88) and in the non-flaring AR ($\sim$ 0.93) show that the flux of both polarities are significantly correlated in both ARs.

The scaling laws obtained from the relations between the size and flux of the patches in the flaring AR ($S \propto F^{0.66}$) and in the non-flaring AR ($S \propto F^{0.64}$), and also, some given samples of time-series of size and flux with highly correlated manner indicate that the size and flux are significantly correlated in both ARs. Moreover, in the previous reports, the relation between the size and flux of patches appeared in the QS can be expressed by $S \propto F^{0.64}$ (Javaherian \textit{et al.} 2017). So, we can say that there would be the universality in the relation between the size and flux of patches all over the solar photosphere. On the other hand, the relations between the flux and lifetime ($F \propto T^{0.48}$), and also, the relation between the size and lifetime ($S \propto T^{0.32}$) demonstrate that the size and lifetime, and also, the flux and lifetime are moderately correlated in the flaring AR; these obtained relations for the non-flaring AR which can be indicated as $F \propto T^{0.37}$ and $S \propto T^{0.23}$ reveal that those parameters are weakly correlated in the flaring AR. As previously reported by Javaherian \textit{et al.} (2017), for the QS, the relation between the flux and lifetime of patches, and also the relation between their size and lifetime were specified as $F \propto T^{0.38}$ and $S \propto T^{0.25}$, respectively. It suggests that the correlated behavior between the flux and lifetime, and also, the relation between the size and lifetime in the QS, is a bit more than that of the non-flaring AR and less than that of the flaring AR. Moreover, the strong correlation between the maximum flux growth rate of the positive polarities and their size, which is obtained as $<dF/dt> \propto S^{0.7}$ in both AR, reveals that the flux growth rate increases with increasing the size of the positive polarities.
\section{Acknowledgements}
The authors acknowledge the \textsf{YAFTA} group, that is, C.E. DeForest, H.J. Hagenaar, D.A. Lamb, C.E. Parnell, and B.T. Welsch, for making \textsf{YAFTA} results publicly available. The authors also thank the SDO/HMI and Solar Monitor science teams for making data publicly available.


\newpage
\begin{thebibliography}{}
\bibitem[{Alipour \& Safari(2015)}]{Alipour2015}
Alipour, N., \& Safari, H. 2015, Statistical Properties of Solar Coronal Bright Points, ApJ, 807, 175-183.
\bibitem[{Aschwanden(2008)}]{Aschwanden(2008)}
Aschwanden, M. J., Stern, R. A., and G{\"u}del, M., 2008, APJ, 672, 659.
\bibitem[{Aschwanden, M. J.}]{}
Aschwanden, M. J., {2011}, {Self-Organized Criticality in Astrophysics}, {Praxis Publishing}, {P84}.
\bibitem[{Aschwanden, M. J.}]{}{Aschwanden, M. J.}, {2015}, {The Astrophysical Journal},{814}, {19}.
\bibitem[{Attie, R., and Innes, D. E.}]{}{Attie, R., and Innes, D. E.}, {2015}, {A\&A}, {574}, {A106}.
\bibitem[{Buehler, D., Lagg, A., van Noort, M., and Solanki, S. K.}]{}{Buehler, D., Lagg, A., van Noort, M., and Solanki, S. K.}, {2019}, {A\&A}, {630}, {A86}.
\bibitem[{Burtseva, O., and Petrie, G.}]{}{Burtseva, O., and Petrie, G.}, {2013}, {Solar Physics}, {283}, {429}.
\bibitem[{Cant\'o, J., Curiel, S., and Mart\'inez-G\'omez, E.}]{}{Cant\'o, J., Curiel, S., and Mart\'inez-G\'omez, E.}
, {2009}, {A\&A}, {501}, {1259}.
\bibitem[{Clauset, A., Shalizi, C. R., and Newman, M. E. J.}]{}{Clauset, A., Shalizi, C. R., and Newman, M. E. J.}, {2009}, {SIAM Review}, {51}, {661}.
\bibitem[{Close, R. M., Parnell, C. E., Longcope, D. W., and Priest, E. R.}]{}{Close, R. M., Parnell, C. E., Longcope, D. W., and Priest, E. R.}, {2005}, {Solar Physics}, {231}, {45}.
\bibitem[{Conlon, P. A., McAteer, R. J., Gallagher, P. T., and Fennell, L.}]{}{Conlon, P. A., McAteer, R. J., Gallagher, P. T., and Fennell, L.}, {2010}, {The Astrophysical Journal}, {722}, {577}.
\bibitem[{DeForest, C. E., Hagenaar, H. J., Lamb, D. A., et al.}]{}{DeForest, C. E., Hagenaar, H. J., Lamb, D. A., et al.}, {2007}, {The Astrophysical Journal}, {666}, {576}.
\bibitem[{de Wijn, A. G., Rutten, R. J., Haverkamp, E. M. W. P., and S{\"u}tterlin, P.}]{}{de Wijn, A. G., Rutten, R. J., Haverkamp, E. M. W. P., and S{\"u}tterlin, P.}, {2005}, {A\&A}, {441}, {1183}.
\bibitem[{de Wijn, A. G., Stenflo, J. O., Solanki, S. K., and Tsuneta, S.}]{}{de Wijn, A. G., Stenflo, J. O., Solanki, S. K., and Tsuneta, S.}, {2009}, {SSRv}, {144}, {275}.
\bibitem[{Everitt, B.}]{}{Everitt, B.}, {2002},{ The Cambridge dictionary of statistics / B.S. Everitt. 2nd ed.}, {Cambridge University Press}, {P346}.
\bibitem[{Farhang, N., Hossein Safari, H., and Wheatland, M. S.}]{}{Farhang, N., Hossein Safari, H., and Wheatland, M. S.}, {2018}, {The Astrophysical Journal}, {859}, {41}.
\bibitem[{Fogel, L., Owens, A., and Walsh, M.}]{}{Fogel, L., Owens, A., and Walsh, M.}, {1966}, {Artificial intelligence through simulated evolution}, {John Wiley \& Sons}, {P73}.
\bibitem[{Gosic, S.}]{}{Gosic, S.}, {2012}, {The Astrophysical Journal}, {749}, {85}.
\bibitem[{Gosi\'c, M., Rubio, L. R. B., del Toro Iniesta, J. C., et al.}]{}{Gosi\'c, M., Rubio, L. R. B., del Toro Iniesta, J. C., et al.}, {2016}, {The Astrophysical Journal}, {820}, {35}.
\bibitem[{Gosi\'c, M., Rubio, L. R. B., Su\'{a}rez, D. O., et al.}]{}{Gosi\'c, M., Rubio, L. R. B., Su\'{a}rez, D. O., et al.}, {2014}, {The Astrophysical Journal}, {797}, {49}.
\bibitem[{Hagenaar, H. J., Schrijver, C. J., Title, A. M., and Shine, R. A.}]{}{Hagenaar, H. J., Schrijver, C. J., Title, A. M., and Shine, R. A.}, {1999}, {The Astrophysical Journal}, {511}, {932}.
\bibitem[{Hagenaar, H. J., Schrijver, C. J., and Title, A. M.}]{}{Hagenaar, H. J., Schrijver, C. J., and Title, A. M.}, {2003}, {The Astrophysical Journal}, {584}, {1107}.
\bibitem[{Honarbakhsh, L., Alipour, N., and Safari, H.}]{}{Honarbakhsh, L., Alipour, N., and Safari, H.}, {2016}, {Solar Physics}, {291}, {941}.
\bibitem[{Javaherian, M., Safari, H., Dadashi, N., and Aschwanden, M. J.}]{}{Javaherian, M., Safari, H., Dadashi, N., and Aschwanden, M. J.}, {2017}, {Solar Physics}, {292}, {164}.
\bibitem[{Jiang, C. W., Feng, X. S., Wu, S. T., and Hu, Q.}]{}{Jiang, C. W., Feng, X. S., Wu, S. T., and Hu, Q.}, {2017}, {Research in Astronomy and Astrophysics}, {17}, {093}.
\bibitem[{Kaithakkal, A. J., Suematsu, Y., Kubo, M., et al.}]{}{Kaithakkal, A. J., Suematsu, Y., Kubo, M., et al.}, {2015}, {The Astrophysical Journal}, {799}, {139}.
\bibitem[{Kaithakkal, A. J., Suematsu, Y., Kubo, M., et al.}]{}{Kaithakkal, A. J., Suematsu, Y., Kubo, M., et al.}, {2013}, {The Astrophysical Journal}, {776}, {122}.
\bibitem[{Kestener, P., Conlon, P. A., Khalil, A., et al.}]{}{Kestener, P., Conlon, P. A., Khalil, A., et al.}, {2010}, {The Astrophysical Journal}, {717}, {995}.
\bibitem[{Liu, H., Liu, C., Wang, J. T. L., and Wang, H.}]{}{Liu, H., Liu, C., Wang, J. T. L., and Wang, H.}, {2019a}, {The Astrophysical Journal}, {877}, {121}.
\bibitem[{Liu, L., Cheng, X., Wang, Y., and Zhou, Z.}]{}{Liu, L., Cheng, X., Wang, Y., and Zhou, Z.}, {2019b}, {The Astrophysical Journal}, {884}, {45}.
\bibitem[{Lorrain, P., Lorrain, F., and Houle, S.}]{}{Lorrain, P., Lorrain, F., and Houle, S.}, {2006}, {Case Study: Solar Magnetic Elements}, {Springer New York}, {P189}.
\bibitem[{Louis, R. E.}]{}{Louis, R. E.}, {2019}, {Journal of Geophysical Research: Space Physics}, {124}, {8255}.
\bibitem[{Mitchell, M.}]{}{Mitchell, M.}, {1996}, {An Introduction to Genetic Algorithms}, {MIT Press}, {P22}.
\bibitem[{Narang, N., Banerjee, D., Chandrashekhar, K., and Pant, V.}]{}{Narang, N., Banerjee, D., Chandrashekhar, K., and Pant, V.}, {2019}, {Solar Physics}, {294}, {40}.
\bibitem[{Noori, M., Javaherian, M., Safari, H., and Nadjari, H}]{}{Noori, M., Javaherian, M., Safari, H., and Nadjari, H}, {2019}, {Advances in Space Research}, {64}, {504}.
\bibitem[{Okunev, O. V., and Kneer, F.}]{}{Okunev, O. V., and Kneer, F.}, {2004}, {A\&A}, {425}, {321}.
\bibitem[{Otsuji, K., Shibata, K., Kitai, R., et al.}]{}{Otsuji, K., Shibata, K., Kitai, R., et al.}, {2007}, {PASJ}, {59}, {S649}.
\bibitem[{Otsuji, K., Kitai, R., Ichimoto, K., and Shibata, K.}]{}{Otsuji, K., Kitai, R., Ichimoto, K., and Shibata, K.}, {2011}, {PASJ}, {63}, {1049}.
\bibitem[{Parnell, C. E.}]{}{Parnell, C. E.}, {2002}, {Monthly Notices of the Royal Astronomical Society}, {335}, {389}.
\bibitem[{Parnell, C. E., DeForest, C. E., Hagenaar, H. J., et al. }]{}{Parnell, C. E., DeForest, C. E., Hagenaar, H. J., et al. }, {2009}, {The Astrophysical Journal}, {698}, {75}.
\bibitem[{P{\'e}rez-Su{\'a}rez, D., Higgins, P. A., Bloomfield, D. S., et al.}]{}{P{\'e}rez-Su{\'a}rez, D., Higgins, P. A., Bloomfield, D. S., et al.}, {2011}, {Automated Solar Feature Detection for Space Weather Applications}, {IGI Global}, {P207}.
\bibitem[{Priest, E.}]{}{Priest, E.}, {2014}, {Magnetohydrodynamics of the Sun}, {Cambridge University Press}, {P25}.
\bibitem[{Scherrer, P. H., Bogart, R. S., Bush, R. I., et al.}]{}{Scherrer, P. H., Bogart, R. S., Bush, R. I., et al.}, {1995}, {Solar Physics}, {162}, {129}.
\bibitem[{Schou, J., Scherrer, P. H., Bush, R. I., et al.}]{}{Schou, J., Scherrer, P. H., Bush, R. I., et al.}, {2012}, {Solar Physics}, {275}, {229}.
\bibitem[{Schrijver, C. J., and Title, A. M.}]{}{Schrijver, C. J., and Title, A. M.}, {2003}, {The Astrophysical Journal}, {597}, {L165}.
\bibitem[{Stenflo, J. O.}]{}{Stenflo, J. O.}, {1973}, {Solar Physics}, {32}, {41}.
\bibitem[{Teh, W. L.}]{}{Teh, W. L.}, {2019}, {Journal of Atmospheric and Solar-Terrestrial Physics}, {188}, {44}.
\bibitem[{Wang, S., Liu, C., Liu, R., et al.}]{}{Wang, S., Liu, C., Liu, R., et al.}, {2012}, {The Astrophysical Journal}, {745}, {L17}.
\bibitem[{Welsch, B. T., and Longcope, D. W.}]{}{Welsch, B. T., and Longcope, D. W.}, {2003}, {The Astrophysical Journal}, {588}, {620}.
\bibitem[{Zender, J. J., Kariyappa, R., Giono, G., et al.}]{}{Zender, J. J., Kariyappa, R., Giono, G., et al.}, {2017}, {A\&A}, {605}, {A41}.
\end{thebibliography}
\end{document}